\documentclass[a4paper,12pt]{article}
\usepackage{caption}
\usepackage{geometry}
\usepackage[pdftex]{graphicx}
\usepackage{graphicx, caption, subcaption}
\usepackage{setspace} 
\usepackage{float}
\usepackage{caption}
\usepackage[authoryear]{natbib}
\usepackage{empheq} 
\usepackage{comment} 
\usepackage{multicol} 
\usepackage{enumitem} 
\usepackage[justification=centering]{caption}
\usepackage[T1]{fontenc}
\usepackage[utf8]{inputenc}
\usepackage{authblk}
\usepackage[table]{xcolor} 
\usepackage{multirow} 
\usepackage{float}
\usepackage{verbatim}
\usepackage{url}
\usepackage{appendix}
\usepackage{ulem}
\usepackage[hang]{footmisc}
\usepackage{multirow}
\usepackage[english]{babel}
\usepackage{textcomp}
\usepackage{amssymb}
\usepackage{multirow}
\usepackage{lipsum}  
\usepackage{booktabs}
\usepackage{adjustbox}
\usepackage{tikz}
\usepackage{xcolor}
\usepackage{rotating}
\usepackage{tablefootnote}
\usepackage[flushleft]{threeparttable}
\usepackage{tabularx}

\usepackage{titlesec}
\usepackage{xr-hyper}

\usepackage{pdflscape}
\usepackage{rotating}
\usepackage{anysize}
\usepackage{titlesec}

\marginsize{2.0 cm}{2.0 cm}{2.0 cm}{2.0 cm}

\usepackage{multirow}
\usepackage{hyperref}

\bibpunct{(}{)}{;}{a}{,}{,}

\singlespace

\begin{document}

%\begin{document}
\title{\textbf{Inter-firm Heterogeneity in Production}}
\author{Michele Battisti \  Valentino Dardanoni \ Stefano Demichelis\thanks{Battisti: University of Palermo,
e-mail: michele.battisti@unipa.it. Dardanoni: University of  Palermo, email: valentino.dardanoni@unipa.it. 
Demichelis: University of Pavia, e-mail: stefano.demichelis@unipv.it.}}
\date{\today}
\maketitle

\begin{abstract}

\noindent  

This paper studies inter-firm heterogeneity in production. Unlike much of the existing research, which primarily addresses heterogeneous production through unobserved fixed effects, our approach also focuses on differences in factors’ output elasticities. Using manufacturing data from Chile, Colombia, and Japan, we apply an innovative Empirical Bayes methodology to estimate heterogeneous Cobb-Douglas production functions. We uncover substantial heterogeneity in both factor neutral productivity and factor elasticities, with a strong negative correlation between them. These findings are consistently observed across datasets and remain robust when using CES and intensive Cobb-Douglas specifications. We show that accounting for these features has significant implications for issues such as markup estimation, firms’ technology adoption, and productivity measurement.\\

\noindent \textbf{J.E.L.}~\textbf{classification codes:}~D24, E23, L00, C11.


\vfill{}

\pagebreak{}
\end{abstract}

%%%%%%%%%%%%%%%%%%%%%%%%%%%%%%%%%%%%%%%%%%%%%%%%%

\pagenumbering{arabic}
\newpage \onehalfspacing \setcounter{page}{2}
%%%%%%%%%%%%%%%%%%%%%%%%%%%%%%%%%%%%%%%%%%%%%%%%%%%%%%%%%%%%%%%%%%%%%%%%%
%%%% Main text entry area:
%%%%%%%%%%%%%%%%%%%%%%%%%%%%%%%%%%%%%%%%%%%%%%%%%%%%%%%%%%%%%%%%%%%%%%%%%

\section{Introduction}\label{s1}

%%%%%%%%%%%%%%%%%%%%%%%%%%%%%%%%%%%%%%

Eighty years ago, \citet{marschak1944random} addressed the issue of interfirm heterogeneity in production techniques, suggesting that production function parameters should be treated as random, due to factors like "\textit{..different technical efficiency, differences in the prices paid or received by various firms (by dropping the assumption of perfect competition); differences in the ability or willingness to choose, or luck in choosing the most profitable combination of resources..}".%
\footnote{\citet{marschak1944random} contributed many seminal insights, including: i) introducing transmission bias in production function estimation—a topic that has since inspired a significant literature (\citealp{olley1996dynamics}, \citealp{levinsohn2003estimating}, \citealp*{ackelberg2015identification}, among others); ii) highlighting identifying challenges of production function estimation under flexible inputs; iii) advancing joint estimation of production functions with profit-maximizing first-order conditions.}

Despite Marschak’s insight, much research has concentrated on unobserved fixed effects, such as managerial quality, with less attention given to heterogeneity in factor elasticities. However, if firms differ solely in factor-neutral productivity (as is common in production studies), a strict dominance order in adopted technologies emerges, which makes it difficult to justify the existence of many competitive firms within the same sector. In contrast, when both factor-neutral productivity and factor returns vary, firms develop distinct strengths and comparative advantages, selecting technologies from an available range of technological options.

This paper introduces a novel Empirical Bayes (EB) approach to estimate the joint distribution of production function parameters.\footnote{EB techniques are becoming popular in economics for addressing parameter heterogeneity. \citet{chen2024empirical} cites over 30 studies in economics employing EB methods in the last five years, and \citet{gu2022} provides a comprehensive introduction to EB methods in the field.}  EB estimation is particularly effective for capturing firm-level heterogeneity in production, allowing us to flexibly model the parameters’ marginal and joint distributions, which allows an unrestricted pattern of association among parameters, firm characteristics, and heteroscedastic errors. Using data from Chilean, Colombian, and Japanese manufacturing firms, we estimate Cobb-Douglas production functions with heterogeneous factor elasticities. Across all samples, we observe substantial interfirm heterogeneity not only in factor-neutral productivity but also in factor returns, with a substantial negative correlation between these two forms of productivity. 

Our estimates of the joint distribution of technology parameters significantly affect productivity measurements. We observe considerable dispersion in both factor-neutral productivity and returns to scale in each sample, with most technological differences occurring within, rather than across, firm size or industry categories. Using \citet{bernard1996comparing}’s Total Technology Productivity (TTP) measure, which considers both productivity types, we find that TTP dispersion is dramatically lower than factor-neutral productivity dispersion: the 90/10 TTP ratios are an order of magnitude smaller than those for TFP, which ignores heterogeneity in factor returns.

Understanding the joint distribution of technology parameters is also crucial for examining issues like market power, markups, and misallocation. Heterogeneity in factors’ parameters affects markup estimates; our findings indicate that our markup estimates show far greater dispersion than Translog estimates due to greater observed heterogeneity.%
\footnote{See \citealp{loecker2012markups}, \citealp{oberfield2021micro}, \citealp{raval2023testing} for recent studies on markup estimation.} Moreover, heterogeneity in production techniques can impact interpretations of market structure. For instance, variations in capital-labor ratios might be attributed to misallocation, although they could result from firms employing different production technologies (\citet{restuccia2017causes}).

Though parameter heterogeneity in production function estimation has long been recognized (at least since \citet{kuh1963capital}) as a source of bias in linear panel models, there have been few direct estimates of heterogeneous production functions. An early approach was proposed by \citet{zellner1966specification}, who suggested estimating Cobb-Douglas parameters as random coefficients. \citet{mairesse1988heterogeneity} applied this approach to Japanese, French, and U.S. firms, finding notable heterogeneity (see \citealp{li2021time} for recent work). Other studies have indirectly addressed heterogeneity by, for example, interacting firm-specific productivity shocks with factor parameters or using polynomial approximations (e.g., \citealp{doraszelski2018measuring}; \citealp{oberfield2021micro}; \citealp*{ackelbergetal2022}; \citealp*{demirer2020production}; \citealp*{li2024identification}). A smaller body of work has explored discrete production technologies through switching or finite mixture regressions (e.g., \citealp{van2003productivity}; \citealp*{battisti2020labor}; \citealp*{kasahara2023identification}).\footnote{These studies generally find that a limited number of techniques (often between two and four) effectively represent firms’ technological range.} 

The remainder of this paper is organized as follows. Section \ref{EB_approach} details our EB approach. Section \ref{Application: Data} introduces the data on Colombian, Chilean, and Japanese manufacturing firms used for estimation. In Section \ref{prodfun}, we estimate a heterogeneous Cobb-Douglas production function and highlight substantial heterogeneity in technology parameters. Section \ref{robustness} presents robustness checks, including CES and intensive Cobb-Douglas estimations. Section \ref{learn} discusses the economic significance of our findings, and Section \ref{conclusion} gives concluding remarks.

 %%%%%%%%%%%%%%%%%%%%%%%%%%%%%%%%%%%%%%%%%%%%%%%%%%%%%%%%%%%%%%%
\section{Empirical Bayes Estimation}\label{EB_approach}

We propose a novel nonparametric EB methodology to estimate the joint distribution of production function parameters.  Suppose we have a sample $(Y_{it}, X_{it})$
where $Y_{it}$ is a measure of output and $X_{it}$ a vector of inputs, \textit{i=1,...,I} denotes the firm and \textit{t=1,....,T} time. In typical panel data applications, $I$ is large, but usually  firms are observed for a short period of time. The production function for $Y_{it}$ can be written as
	\begin{equation}\label{eq:firsteq}
	Y_{it} = h \left (X_{it}; \psi_{i} \right ) + \epsilon_{it}
	\end{equation} 
where  $\psi_i$,  $i=1,...,I$, is a vector of heterogeneous firm specific technology parameters,  and  $\epsilon_{it}$ is an idiosyncratic normally distributed i.i.d. error with zero mean and (heteroschedastic) standard deviation $s_i$. Notice that while we are assuming normality of
$\epsilon$, we are not making any specific distribution assumption on the  distribution of the technology parameter $\psi$ and its joint distribution with $\epsilon$.  

The set of parameters is defined by a vector $\theta = (\psi,s)$, with individual firm parameters $\theta_i = (\psi_i,s_i)$,  $i=1,...,I$. To guarantee identification of the heterogeneous  parameter $ \psi_{i}$, we assume that the function $h \left (X_{it}; \psi_{i} \right )$ in \eqref{eq:firsteq} is injective for all $X$. Notice that a necessary condition for identifiability is that the number $T$ of observations for each firm is greater than the size of 
$\psi$,  a significant constraint to the application of these methods with short panels.

We define a finite number of unobserved firm \textit{types} by discretizing $\theta$ into finite grids of sizes $M^1, M^2,...., M^{J}$, arranged lexicographically, where $J$ denotes size of $\theta$. There are $\prod_{j=1}^J  M^j  = Q$  types: an (unknown) firm type $q$ ( $q=1,...,Q$ )  is a tuple of discretized parameters 
$\theta$. The discretized joint distribution of the parameters is a $Q \times 1$ vector, described by a probability vector $\pi$. Notice that the joint distribution of $\theta$ is unrestricted (a discrete nonparametric approximation of the true distribution).%
\footnote{Typically grids are built with equally spaced points in intervals which contain most of the parameters' density.}  
For any firm \textit{i}, conditional on being of type  $q$, having inputs $X_{it}$, and error variance $s_i^2$, the conditional  density of $Y_{i}$ is
\begin{equation}\label{eq:density}
f_{iq} = f(Y_{i} \mid q, X_{it}, s_i) = \prod_{t=1}^T \frac{1}{s_i} \varphi \left( \frac{Y_{it} - h(X_{it};q)}{s_i} \right).
\end{equation}
We collect $f_{iq}$ into the $I \times Q$ matrix  $F$. 

Given a prior distribution, say, $\pi^0$, by Bayes Theorem we find  the posterior probability that firm $i$ is of type $q$
	\begin{equation*}
h_{iq} = \frac{f_{iq} \times \pi^0_q }{ \sum_v f_{iv} \times \pi^0_v}.  % \ q = 1,\dots,Q.
	\end{equation*}
Repeating for all firms and all types, we collect $h_{iq}$ in the $I \times Q$  matrix  $H$. 
$H$ is a function of $F$ and $\pi^0$, and
the $i$th row of $H$  gives the posterior probability of firm $i$ being a given unobserved type. If we take the mean of each column of $H$, we get an estimate of the posterior distribution, say $\hat{\pi}$, of the discretized joint distribution of the parameters $\theta$.

Any Bayesian method requires the choice of an appropriate prior distribution ($\pi^0$ in our setting). EB estimation typically uses the available data to specify a (parametric or nonparametric) prior. It is straightforward to see that, in this setting, the prior and posterior distributions are multinomial, with a common support on a fixed set of states.
Using an approach suggested by \citet{dardanoni2023}, we propose that the choice of the prior is guided by two rational expectations conditions:  (1) the choice of the prior is {\textit coherent} with the posterior (that is $\pi^0 = \hat{\pi}$), which conveys the idea that, if given the evidence brought by the data, the posterior distribution agrees with the prior, there is no reason the change the prior;  (2) the prior is \textit{stable}, in the sense that if it puts zero weight on a parameter configuration, Bayesian updating of nearby priors decreases its weight.

Formally, coherence and stability imply that estimating the posterior distribution $\hat{\pi}$ is equivalent to finding a stable fixed point in a nonlinear system of equations from a probability simplex into itself.   \citet{dardanoni2023} show that a coherent and stable fixed point (say   
$\pi^*$) exists and is unique.%
\footnote{While existence follows immediately from Brouwer Theorem, the fixed point may not be necessarily unique: any degenerate probability vector is a fixed point in this context.
Requiring that $\pi^0$ has full support would be inappropriate since, when there are few observations compared with the size of $\pi^0$, we would expect many zeros. The stability property allows us to sort actual zeros from artifacts.}
 Furthermore, they show that the number of strictly positive points in $\pi^*$ is  less or equal than the number of firms $I$, and  $\pi^*$ is the unique Maximum Likelihood estimate of $\pi$.%
 \footnote{Notice the relationship with Nonparametric Maximum Likelihood Estimation by \citet{kiefer1956} and \citet{mizera2014}.}
 
 The unique fixed point $\pi^*$ can be found by an iterative procedure,\footnote{Formally this is equivalent to using the E step in the EM algorithm.}  which is guaranteed to converge to the unique global maximum of the Log Likelihood.
 From the estimated $\pi^*$, by appropriate marginalization, we can directly derive statistics such as means, standard deviations, and correlations of the joint distribution of  the production function parameters. Furthermore, given $\pi^*$, using the rows of the matrix of posterior probabilities $H(F,\pi^*)$, we  immediately  get an estimate of the production function parameters for each individual  firm $i=1,\dots,I$ by taking their expected values.

 %%%%%%%%%%%%%%%%%%%%%%%%%%%%%%%%%%%%%%%%%%%%%%%%%%%%%%%%%%%%%%%
\section{Data}\label{Application: Data}

In this study, we use three datasets: plant censuses for Chilean and Colombian firms (popular choices in this field, see e.g. \citealp{levinsohn2003estimating}, \citealp{gandhi2020identification} \citealp{raval2023testing}, among others) and Orbis Bureau Van Dijk database for Japan, which offers the highest Orbis coverage among OECD countries. The Japanese data was extracted in July 2022.

Our analysis covers countries with different development levels, different periods of time, and varying numbers of firms per country. 
%We then compare our results with those obtained using more traditional techniques (e.g., translog) to ensure that any observed heterogeneity within a country is not merely a result of idiosyncratic latent characteristics. 
For output ($Y$), we use value added, for labor ($L$), the number of workers, for capital ($K$), fixed assets. For ease of notation, we denote the natural logs of $Y$, $K$, and $L$ as $y$, $k$, and $l$, respectively.

Since attrition may introduce additional heterogeneity, we include only firms operating throughout the entire period, always in the same sector, without addressing the selection issue highlighted by \citealp{olley1996dynamics}. In addition to this, we need data on intermediates to estimate the Translog production function for comparison, and data on wage share to compute labor markups, so we exclude missing data also on these two variables. 

The final balanced samples for the entire period, include 1,535 Colombian firms between 1978 and 1989 (18,420 observations) and 1,096 Chilean firms between 1986 and 1996 (12,056 observations). For Japan, we adopt a conservative approach to reduce outlier heterogeneity. We apply the \citet*{billor2000bacon} algorithm to exclude 10\% of univariate outliers from the marginal distributions of $(Y, K, L)$ and multivariate outliers from their joint distribution\footnote{For example, \citet{raval2023testing} excludes 2\% of observations to avoid extreme values. Removing outliers can be particularly important in the presence of noisy capital measurements, which is a known issue in the Orbis dataset, as discussed by \citet{gopinath2017capital}.}. The final sample consists of 39,116 observations from 5,588 firms between 2013 and 2019 (we stop at 2019 to avoid using budget data may be influenced by the Covid period). Table \ref{tab_stat} below reports the sample statistics for the three working samples with variables in logs. 

\vspace{0.5cm}
	
	\begin{table}[hbt!]
%		\vspace{-8mm}
		\centering
%		\resizebox{\columnwidth}{!}{
		\caption{Working sample statistics \label{tab:pw_distribution}}
	%	\begin{adjustbox}{width=0.6\textwidth}
	%		\begin{threeparttable}
			\small{
				\begin{tabular}{l|ccc|ccc|ccc}
					\toprule
					&\multicolumn{3}{c}{Chile}&\multicolumn{3}{c}{Colombia}&\multicolumn{3}{c}{Japan}\\
					\hline
					& \textit{y} & \textit{k}&\textit{l} &\textit{y}&\textit{k}&\textit{l}&\textit{y}&\textit{k}&\textit{l} \\
					\midrule
%					$1^{st}$ quartile   & 6.95  & 7.36&3.00&&&&7.31&7.79&3.33 \\
%					%\\
					Median   & 10.96  &8.81 &3.79&9.63&7.83&3.93&7.75 &8.36&3.74\\
%					\\
					Mean   & 11.16  &9.00 &4.02&9.88&8.01&4.09& 7.85&8.36&3.84\\
%					\\
%					$3^{rd}$ quartile   & 8.67  & 9.44&4.61&&&&9.08&9.80&4.91 \\
%					%\\
					Standard deviation  & 1.86 &2.02 &1.10 &1.77&1.82&1.16&1.14&1.36 &1.00 \\
%&\multicolumn{3}{r}{|}	&\multicolumn{3}{r}{|}&\multicolumn{3}{r}{|}				\\
					\hline
					Obs&\multicolumn{3}{c}{12056}&\multicolumn{3}{c}{18420}&\multicolumn{3}{c}{39116}\\
				Firms&\multicolumn{3}{c}{1096}&\multicolumn{3}{c}{1535}&\multicolumn{3}{c}{5588}\\
					\hline\\
					\bottomrule
				\end{tabular}\medskip{}
%					\footnotesize
%					\vspace{-3mm}
%			\end{threeparttable}
%		\end{adjustbox}
}
		\label{tab_stat}
	\end{table}
%

%As expected, the standard deviations of output are lower in Japan, confirming insights from the misallocation literature. Additionally, the means are slightly higher than the medians, likely due to the presence of large firms on the right side of the distribution. The variables in their working samples tend to have higher means (for instance in Japan means are increased by 5\% for value added, 4\% for capital stock and 6\% for labor due to the natural selection process of \citealp{jovanovic1982}) and lower standard deviations (standard deviations are reduced by 17\% for value added, 12\% for capital stock and 7\% for labor). 

%%%%%%%%%%%%%%%%%%%%%%%%%%%%%%%%%%%%%%%%%

\section{Heterogeneous Cobb-Douglas production function}\label{prodfun}

The Cobb-Douglas (hereon CD) production function (\citealp{cobb1928theory}) may be defined as
\begin{equation}\label{eq:cd}
	y_{it} = \alpha_{it}+ \beta_i k_{it}+\gamma_i l_{it} + \xi_{it} 
\end{equation}
where $\alpha$ is a factor neutral technology term (such as TFP, Hicksian-neutral productivity, and managerial ability),
$\beta$ and $\gamma$ output elasticities, and $\xi_{it}$ a zero mean error.
% We may think of the set ($\alpha$, $\beta$, $\gamma$) as Total Technological Productivity (TTP) (\citealp{bernard1996comparing}).
 
 Notice that we are assuming time and firm heterogeneity in $\alpha$ but firm heterogeneity alone in $\beta$ and $\gamma$, since some components of \textit{$\alpha$}, as for instance managerial quality, may change even when factor endowments remain unchanged. Our modeling strategy recognizes that intercepts may be correlated with inputs changes, possibly inducing endogeneity bias (\citealp{olley1996dynamics}), but time invariance in output elasticities implies that, after a firm has determined its factor endowments, in the short run the embodied technology (\citealp{hulten1992growth}) may only change minimally over time.\footnote{The flexibility of our approach allows for the incorporation of time-varying effects also in factor elasticities if sufficient data were available. Increasing the number of estimated parameters may require longer panels and a very large number of firms for accurate estimation.} 
We capture the time-varying heterogeneity in $\alpha_{it}$ through individual coefficients 
($\alpha^0_i$, $\alpha^1_i$, $\alpha^2_i$)
\begin{equation}\label{eq:6}
	\alpha_{it}= \alpha^0_i + \alpha^1_{i}t + \alpha^2_{i}t^2 + \eta_{it}
\end{equation}
with $\eta_{it}$ as a zero mean error.
Our modeling of the productivity term $\alpha_{it}$ is similar to that of \citet*{ackelbergetal2022}, with a key difference. In our case, time-varying heterogeneity is captured through heterogeneous 
$(\alpha^1,\alpha^2)$ coefficients, whereas in their setting, $\alpha_{it}$ is split into a homogeneous process linked to time and an idiosyncratic third order Markov process, capturing individual heterogeneity.\footnote{The setting in \eqref{eq:6} is thought to deal with the transmission bias issue of \citet{marschak1944random}, pioneered in the last three decades by \citet{olley1996dynamics}, under the assumption that the firm specific second order polynomial fully captures 
heterogeneous productivity dynamics. Note that this implies the manager is aware of the firm's productivity evolution over time (as estimated by our quadratic trend setting) and chooses inputs accordingly. In other words, we assume that residual innovations to productivity are observed after input decisions are made. As noted earlier, this approach may be more reasonable over shorter time spans.}

The final estimating equation is
\begin{equation}\label{eq:final}
	y_{it} = \alpha^0_i + \alpha^1_{i}t + \alpha^2_{i}t^2 + \beta_i k_{it}+\gamma_i l_{it} + \phi_{it} 
\end{equation}
with $\phi_{it} \sim N(0,s_i)$, thus allowing for heteroschedasticity. Comparing equations \ref{eq:cd} and \ref{eq:final}, notice that
$\phi_{it}$ plays the role of an independent and idiosyncratic normal error as in equation \ref{eq:firsteq}, while
these restrictions do not hold for $\xi_{it}$ in equation \ref{eq:cd}.  

Henceforth we will denote the firm specific time averaged idiosyncratic factor neutral productivity, which sometimes will be referred as the ``intercept'', as $\bar{\alpha}_i=\frac{1}{T}\sum^T_{t=1} (\alpha^0_i + \alpha^1_{i}t + \alpha^2_{i}t^2)$, where $T=11, 12, 7$
 respectively for Chile, Colombia and Japan.

%%%%%%%%%%%%%%%%%%%%%%%%%%%%%%%%%%%%%%%%%%%%%%%%%%%%%%%%%%

\subsection{Results}\label{Results}

Before illustrating our EB estimation results, we exploit our panel data structure to estimate firms' parameters individually for each firm \textit{i} by OLS,
(as in \citet{mairesse1988heterogeneity}), after checking that each firm's regression matrix is full rank. Clearly  with short panels this may be \textit{very} inefficient. In all the samples, not only the means of the estimated factor elasticities may be unrealistic, but as shown in Figure \ref{CD_OLS} in the Appendix, there is an enormous dispersion in estimated parameters, suggesting a need for some form of regularization. The  EB approach, by ``borrowing strength'' from the whole sample of firms, improves the efficiency of the compound estimation of the parameters' joint distribution.

Turning to our EB approach, to estimate the rational expectation prior $\pi^*$ we need to find an appropriate discretization of the estimating parameters ($\alpha^0, \alpha^1, \alpha^2, \beta, \gamma, s$). We discretize $\alpha^0$, $\beta$, and $\gamma$  with 15 
and $\alpha^1$, $\alpha^2$ and $s$ in 6 equally spaced  points, using non negative grids for $\beta$, and $\gamma$.  We end up with $15^3 6^3=729,000$ parameters' configurations which nonparametrically approximate the true joint distribution. 

We combine equations \eqref{eq:density} and \eqref{eq:final} to find the density matrix $F$ which collects the conditional densities for all firms. Having obtained $F$ for the three samples, we apply an iterative algorithm (starting with a uniform distribution)  to find the fixed point $\pi^*$.
As expected, in each sample  the number of non zero points in the support is lower than the number of firms.

 From $\pi^*$, we directly get the parameters' means, which are shown in Table \ref{tab_meanCD} below. Chile has an average scale elasticity greater than one, Colombia’s is approximately one, and Japan’s is less than one. The labor output elasticities vary significantly across the three samples, while the capital output elasticities show more similarity.
 
 We compare our estimates of the mean factor elasticities’ parameters with standard estimators. Table \ref{tab:hom_est} in the Appendix shows that empirical Bayes (EB) estimates are largely consistent with those obtained using the Ackerberg, Caves, and Frazer (ACF) Translog production function approach (\citealp{ackelberg2015identification}) and are slightly smaller than the OLS estimates across all samples.
The ACF Translog production function addresses endogeneity in productivity shocks, as described by \citet{ackelberg2015identification}. For our estimation, we use the ACF Stata prodest routine developed by \citet{rovigatti2018theory}.
  
\vspace{0.25cm}
 %TABELLA 2 ESPANDERE NEL TESTO

\begin{table}[hbt]
		\vspace{-2mm}
		\centering
		 \caption{Average value of EB CD parameters\label{tab:pw}}
		\begin{adjustbox}{width=0.7\textwidth}
			\begin{threeparttable}
				\begin{tabular}{lcccccc}
					\toprule
					& $\alpha^0$ & $\beta$& $\gamma$  &$\alpha^1$&$\alpha^2$&$s$\\
					\midrule
					\multirow { 2}{*}{Chile} & 4.035  & 0.425 & 0.789 & 0.025 & -0.002 & 0.408 \\
					& (0.084) & (0.009) & (0.016) & (0.005) & (0.000) & (0.005)\\
					%\\
					\hline
					%Observations &  \multicolumn{6}{c}{12056}\\
					%Firms&  \multicolumn{6}{c}{1096}\\
					\hline
\multirow { 2}{*}{Colombia} & 3.996  & 0.407 & 0.668 & -0.055 & 0.004 & 0.373 \\
					& (0.065) & (0.007) & (0.011) & (0.003) & (0.000) & (0.004)\\
					%\\
					\hline
					%Observations &  \multicolumn{6}{c}{18420}\\
					%Firms&  \multicolumn{6}{c}{1535}\\
					\hline
\multirow { 2}{*}{Japan} & 3.491  & 0.329 & 0.387 & 0.022 & -0.001 & 0.184 \\
					& (0.023) & (0.003) & (0.003) & (0.002) & (0.000) & (0.001)\\
					%\\
					\hline
					%Observations &  \multicolumn{6}{c}{39116}\\
					%Firms&  \multicolumn{6}{c}{5588}\\
					\bottomrule
				\end{tabular}\medskip{}
				\begin{tablenotes}
					%\singlespacing
					%\vspace{-5mm}
					\item \textit{Notes}: Standard errors in parentheses.
					{\footnotesize\par}
				\end{tablenotes}
					\footnotesize
					\vspace{-3mm}
			\end{threeparttable}
		\end{adjustbox}
		\label{tab_meanCD}
	\end{table}
	
\vspace{0.05cm}

%%%%%%%%%%%%%%%%%%%%%%%%%%%%%%%%%%%%
	
\subsection{Parameters heterogeneity}\label{ss_CD_het}

\begin{figure}[H]
\centering
  \includegraphics[width=.83\textwidth]{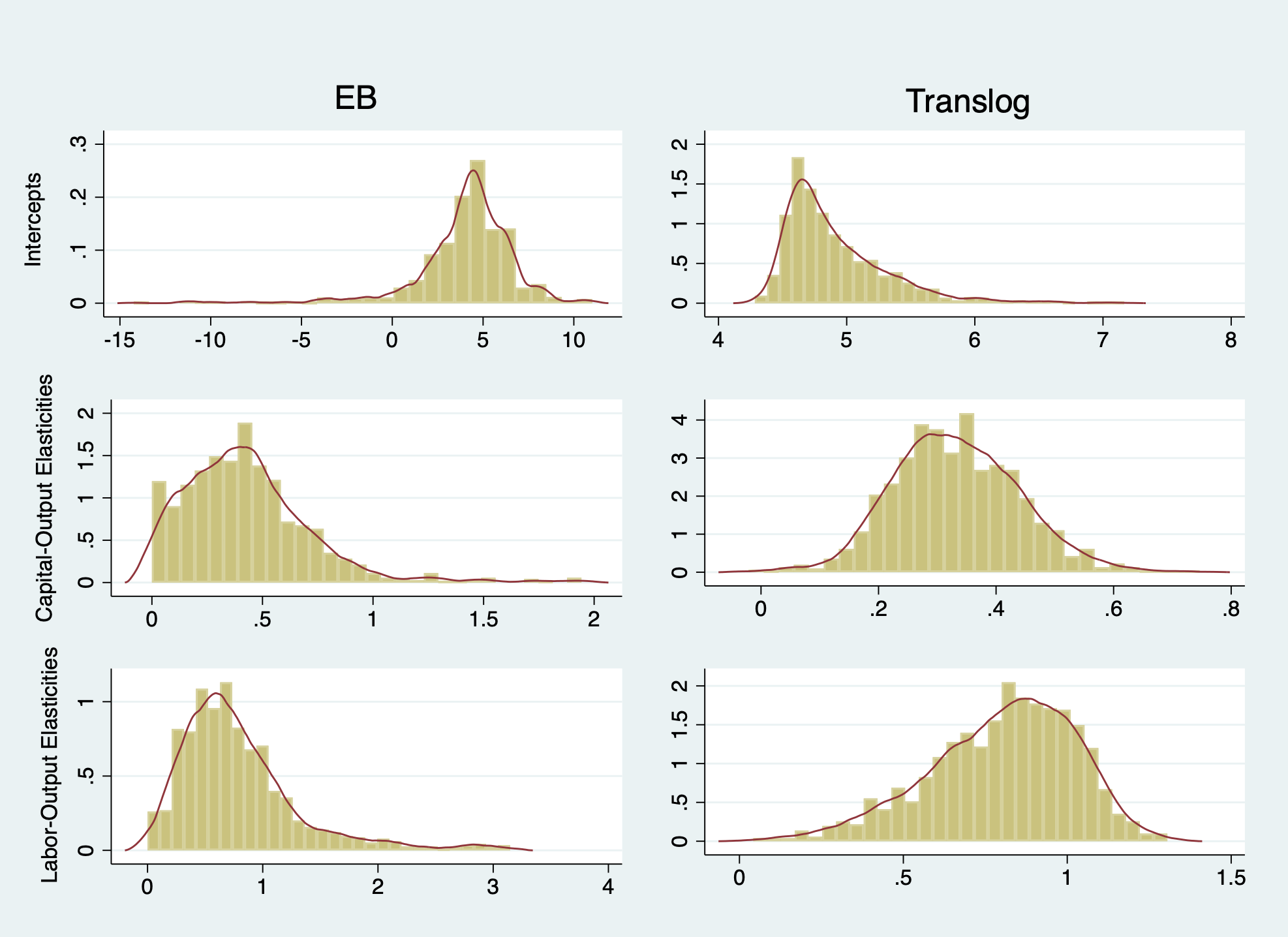}
  \caption{Distributions EB vs Translog Chile}\label{CD_CHL}
\end{figure}

\begin{figure}[H]
\centering
  \includegraphics[width=.83\textwidth]{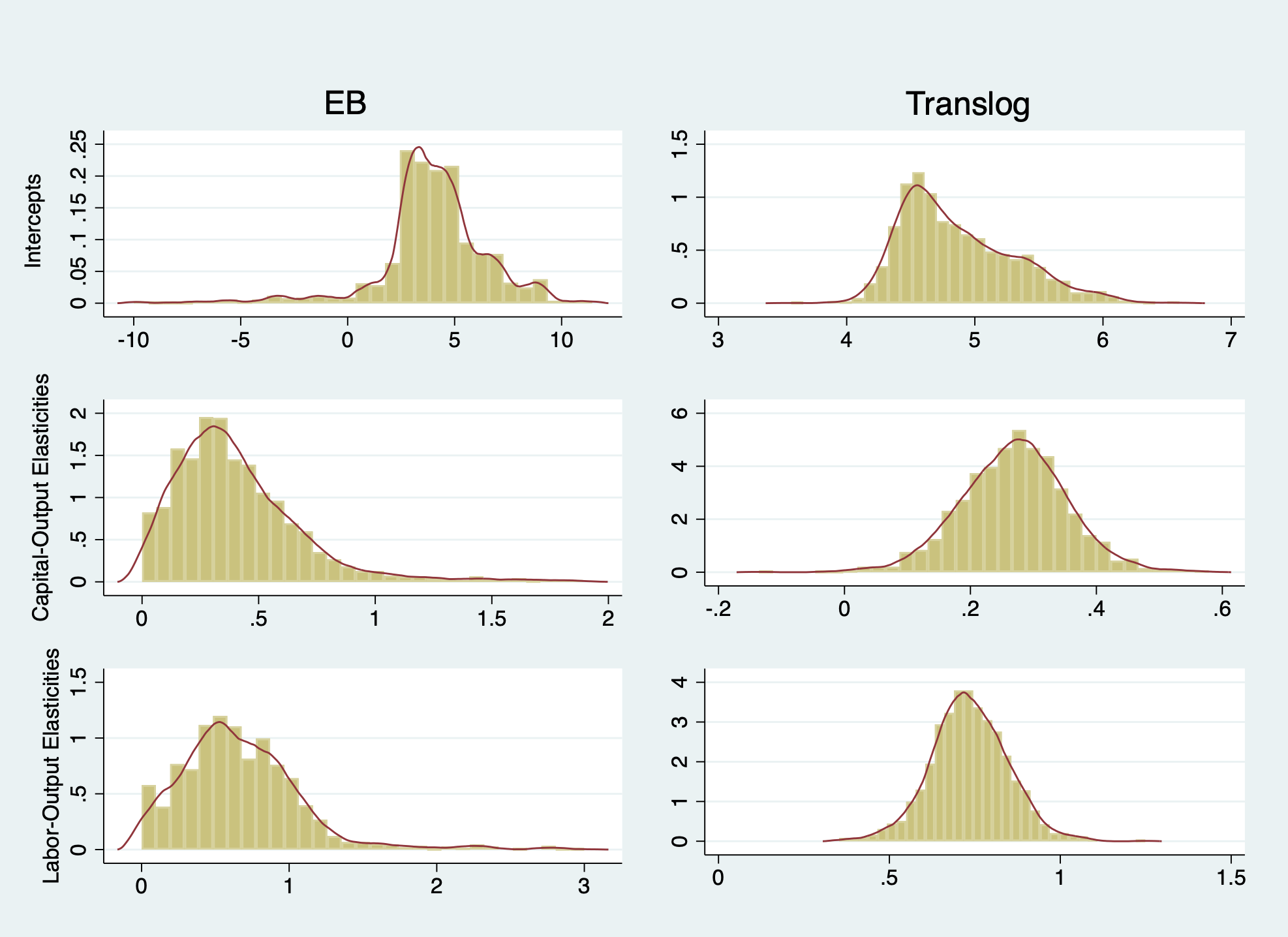}
  \caption{Distributions EB vs Translog Colombia}\label{CD_COL}
\end{figure}

\begin{figure}[H]
\centering
  \includegraphics[width=.83\textwidth]{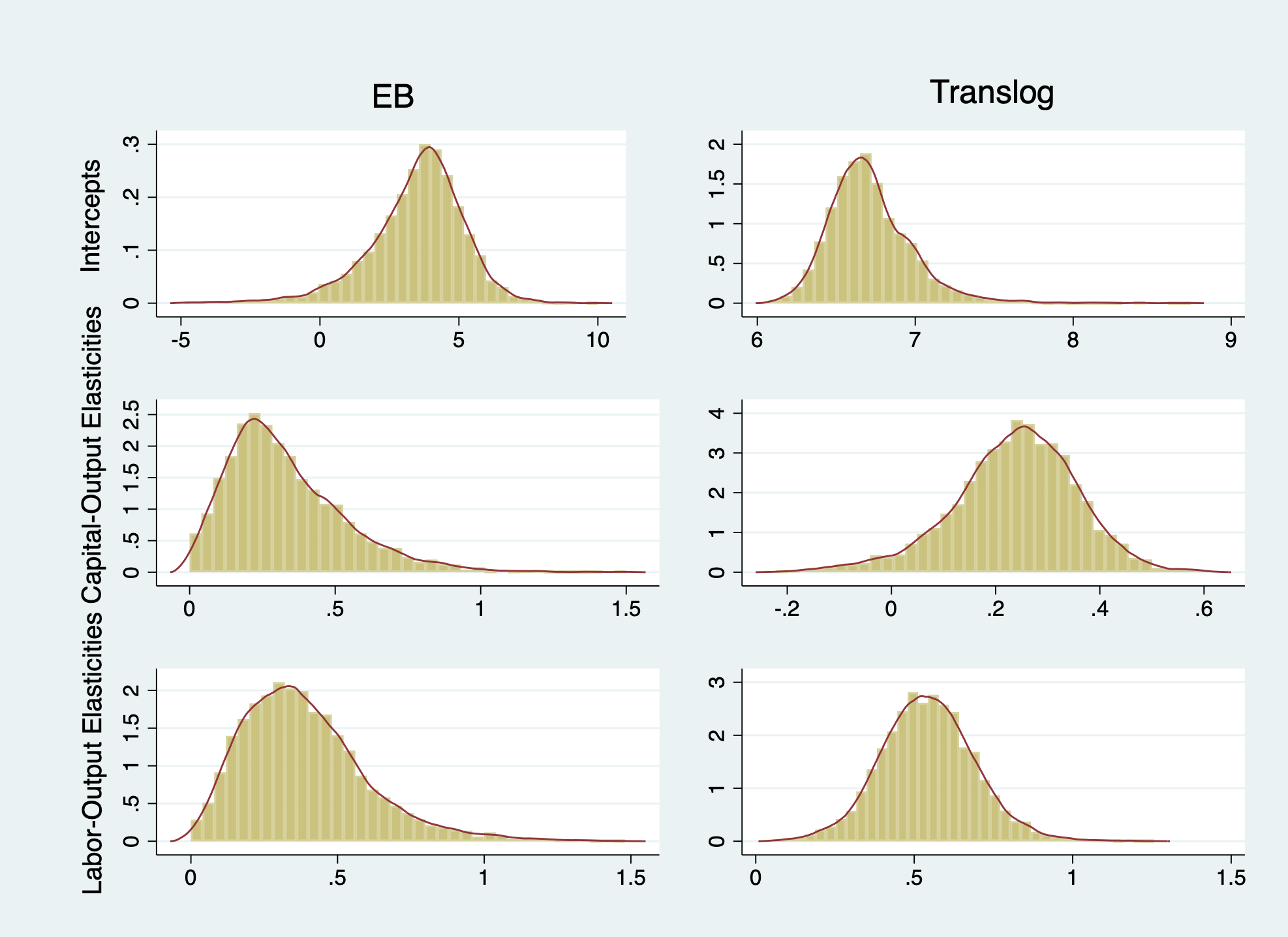}
  \caption{Distributions EB vs Translog Japan}\label{CD_JAP}
\end{figure}

Table \ref{tab_dispersion_CD} shows some dispersion metrics for our EB estimates and for the ACF Translog. The standard deviation of the intercept is approximately six times greater, that of the capital-output elasticity about twice as large, and that of the labor-output elasticity between two and three times larger in the EB estimates compared to the Translog estimates. Similarly, differences between the 90th and 10th percentiles also indicate substantial heterogeneity, and more so for the EB estimates compared to the Translog ones. Looking at Figures \ref{CD_CHL}, \ref{CD_COL}, \ref{CD_JAP}, the EB estimates generally show a more dispersed and asymmetric distribution with pronounced long tails.

 \begin{table}[hbt]
		%\vspace{-2mm}
		\centering
		 \caption{Comparison of elasticities: EB and Translog\label{tab:pw}}
		\begin{adjustbox}{width=0.7\textwidth}
			\begin{threeparttable}
				\begin{tabular}{lcccccc}
					\toprule

\multicolumn{7}{c}{Chile}	\\
\hline					
&\multicolumn{2}{c}{Intercept}&	\multicolumn{2}{c}	{Capital-output elasticity}&\multicolumn{2}{c}	{Labor-output elasticity}\\	
\hline
	&Translog&	EB	&Translog	&EB	&Translog&	EB\\
%Mean&	4.391	&4.035	&0.336&	0.425&	0.813&	0.790\\
Median&	4.270	&4.356	&0.333&	0.396&	0.836&	0.684\\
Std dev	&0.375	&2.765	&0.107	&0.289&	0.245&	0.525\\
P90/P10&	1.213&	5.024	&2.264&	8.123	&2.345&5.408	\\
%P90/P50&	1.149&	1.503	&1.409&	1.911	&1.333&	2.078\\
\hline\\
	\multicolumn{7}{c}{Colombia}	\\
	\hline					
&\multicolumn{2}{c}	{Intercept}&	\multicolumn{2}{c}{Capital-output elasticity}&\multicolumn{2}{c}	{Labor-output elasticity}\\	
	\hline
	&Translog&	EB	&Translog	&EB	&Translog&	EB\\
%Mean&	4.792	&3.996&	0.262&	0.407&	0.728&	0.667\\
Median&	4.704	&3.993	&0.264&	0.354&	0.726&	0.621\\
Std dev&	0.470&	2.555&	0.080&	0.279&	0.115	&0.415\\
P90/P10&	1.278&	3.832	&2.203&	6.206	&1.489&	5.357\\
%P90/P50&	1.161&	1.735&	1.361&	2.055&	1.208	&1.790\\
\hline\\
	\multicolumn{7}{c}{Japan}		\\	
	\hline			
&\multicolumn{2}{c}	{Intercept}&	\multicolumn{2}{c}	{Capital-output elasticity}&\multicolumn{2}{c}{Labor-output elasticity}\\	
	\hline
	&Translog&	EB	&Translog	&EB	&Translog&	EB\\
%Mean&	6.714	&3.492&	0.237	&0.329&	0.543	&0.387\\
Median&	6.681	&3.701	&0.245	&0.290	&0.540	&0.358\\
Std dev&	0.265&	1.700&	0.119	&0.203&	0.149	&0.213\\
P90/P10&	1.096&	4.011	&4.541&	5.607	&2.024&	4.654\\
%P90/P50	&1.054	&1.454&	1.548&	2.078&	1.351	&1.871\\
					\bottomrule
				\end{tabular}\medskip{}
					\footnotesize
					\vspace{-3mm}
			\end{threeparttable}
		\end{adjustbox}
		\label{tab_dispersion_CD}
	\end{table}
	 
 Notice also that while EB estimates capture the heterogeneity in structural technological parameters, the Translog production function, as remarked  by \citet{raval2023testing},  "..\textit{allows the output elasticities to vary based upon inputs, but they remain a deterministic function of production parameters and inputs with no error term. Thus, the Translog estimated output elasticities cannot capture the full degree of heterogeneity in input shares..}".
 In other words, the Translog production function assumes uniform technological parameters, and differences in factor elasticities arise because firms use varying levels of inputs.

 \vspace{0.5cm}

%, and their correlations. 

%\vspace{1cm}

\begin{table}[hbt]
		%\vspace{-2mm}
		\centering
		\caption{CD: firm specific parameters correlation \label{tab:pw_distribution2}}
		\begin{adjustbox}{width=0.3\textwidth}
			\begin{threeparttable}
				\begin{tabular}{lcccc}
					\toprule
					\multicolumn{5}{c}{Chile}\\
					\hline
					&&$\beta$ &$\gamma$&$\beta+\gamma$\\
					\hline
					&$\alpha^0$  &\multicolumn{1}{|c}{-0.661}&-0.456&-0.828\\
&$\beta$&\multicolumn{1}{|c}{}   & -0.292&0.262\\
&$\gamma$&\multicolumn{1}{|c}  {}    &&0.847\\
\hline
\multicolumn{5}{c}{Colombia}\\
					\hline
					&&$\beta$ &$\gamma$&$\beta+\gamma$\\
					\hline
					&$\alpha^0$  &\multicolumn{1}{|c}{-0.703}&-0.488&-0.887\\
&$\beta$&\multicolumn{1}{|c}{}   & -0.207&0.430\\
&$\gamma$&\multicolumn{1}{|c}  {}    &&0.795\\
\hline

\multicolumn{5}{c}{Japan}\\
					\hline
					&&$\beta$ &$\gamma$&$\beta+\gamma$\\
					\hline
					&$\alpha^0$  &\multicolumn{1}{|c}{-0.831}&-0.130&-0.819\\
&$\beta$&\multicolumn{1}{|c}{}   & -0.337&0.548\\
&$\gamma$&\multicolumn{1}{|c}  {}    &&0.604\\
\hline

%&&\multicolumn{1}{|c}{}&\\
					\bottomrule
				\end{tabular}\medskip{}
				
				%\begin{tablenotes}
					\footnotesize
					\vspace{-3mm}
			\end{threeparttable}
		\end{adjustbox}
		\label{tab_CorrCD}
	\end{table}

Finally, we consistently find very strong negative correlations between the intercepts and the factors' output elasticities in Table \ref{tab_CorrCD} in all samples.%
\footnote{A similar apparent puzzle was found and discussed in \citet{mairesse1988heterogeneity} and \citet{li2021time}. \citet{li2021time} suggests the possibility that this is due to latent omitted factors affecting production.} 
 We discuss possible interpretations for this rather surprising  result in Subsection \ref{ss:neg} below.

%%%%%%%%%%%%%%%%%%%%%%%%%%%%%%%%%%%%%%%%%%%%%%%%%%%%%%%%%%%

\section{Robustness and further analysis}\label{robustness}

%%%%%%%%%%%%%%%%%%%%%%%%%%%%%%%%%%%%%%%%%%%%%%%%%%%%%%%%%%%

\subsection{Simulation}

To check the validity of our EB procedure, we perform a simulation using a CD production function
\begin{equation}\label{eq_sim}
	y_{it} = \alpha_i + \beta_{i} k_{it} + \gamma_i l_{it} + \epsilon_{it}, \quad \epsilon_{it} \sim N(0,\sigma_i)
\end{equation}
where, for simplicity, we set the dynamic coefficients $\alpha_1$ and $\alpha_2$ to zero.
We use the same structure as our Japanese dataset, using the sample values of $(k_{it}, l_{it})$ with $N=5588$ and $T=7$. To generate the heterogeneous coefficients $(\alpha_i, \beta_i, \gamma_i, \sigma_i)_{i=1,\dots,N}$, we draw from a multivariate normal distribution reflecting the joint distribution of  $(\hat{\alpha}_i, \hat{\beta}_i, \hat{\gamma}_i, \hat{\sigma}_i, k_i, l_i)$ (the estimated individual coefficients and the mean values of capital and labor for the \textit{N} firms). Overall, this procedure generates realistic heterogeneous parameters with plausible correlations to the firms' input values, providing a non-discretionary structure to the simulation.

We generate $b=1,\dots,100$ samples of firm-specific parameters $(\alpha^b_i, \beta^b_i, \gamma^b_i, \sigma^b_i)$, and for each sample $b$, we generate firms' output as:
\[
y_{it}^b = \alpha_i^{b} + \beta_{i}^b k_{it} + \gamma_{i}^b l_{it} + e_{it}^b
\]
where $e_{it}^b$ is a draw from $N(0,\sigma^b_i)$. For each sample $b$, we estimate $(\hat{\alpha}^b_i, \hat{\beta}^b_i, \hat{\gamma}^b_i, \hat{\sigma}^b_i)$ using our EB method, and compare the estimated parameters with the ``true'' values used in the DGP.

Table \ref{tab:EB_simulation} shows the Bias and MSE of the mean and standard deviation of the estimated parameters. We observe that the means of all individual parameters are estimated quite precisely, but their dispersion is underestimated. This is expected, as it is well known in the EB literature that the variance of the expected value of the estimated individual parameters tends to underestimate the true variance. (\citealp{kass1989approximate}). 

\begin{table}[hbt]
    \centering
    \caption{EB Simulation\label{tab:EB_simulation}}
    \begin{adjustbox}{width=0.35\textwidth}
        \begin{threeparttable}
            \begin{tabular}{lccc}
                \toprule
                \multicolumn{4}{c}{BIAS} \\
                & $\alpha$ &  $\beta$ & $\gamma$ \\
                \hline
                mean  & 0.031  &  -0.002 &  -0.004  \\
                std & 0.488  &  0.062  &  0.059   \\
                %\\
                \hline
                \multicolumn{4}{c}{MSE} \\
                & $\alpha$ &  $\beta$ & $\gamma$  \\
                \hline
                mean  & 0.002  &  $< 10^{-4}$  &  $< 10^{-4}$   \\
                std & 0.237 &   0.004  &  0.004   \\
                \bottomrule
            \end{tabular}
        \end{threeparttable}
    \end{adjustbox}
\end{table}

%%%%%%%%%%%%%%%%%%%%%%%%%%%%%%%%%%%%%%%%%%%%%%%%%%%%%%%%%%%

\subsection{CES}

The CD production function implicitly assumes that the elasticity of substitution between capital and labor is unitary. 
As a further robustness check, and to show the general applicability of our EB approach, we estimate a well known generalization of the CD,  the CES  production function proposed by \citet{arrow1961capital}. Details of the EB estimation CES are in section \ref{Het_CES} in the Appendix. We see that even in presence of large and heterogenous elasticity of substitution among inputs, in all three samples there is substantial consistency between CD and CES estimation with respect to comparable estimates: we see a remarkably similar distributions of both factor neutral productivity and returns to scale, and a similar strong negative correlation.  

%%%%%%%%%%%%%%%%%%%%%%%%%%%%%%%%%%%%%%%%%%%%%%%%%%%%%%%%%%%

\subsection{Intensive CD production function}\label{intensive}

For robustness we also estimate a simple heterogeneous CD intensive production function
\[
\ln Y_{it} = a_i + b_i  \ln X_{it} + \epsilon_{it}
\]
in our three samples, with $Y$ being output for worker, $X$ capital for worker, and $\epsilon_{it} \sim N(0,\sigma_i)$. The EB estimates are reported in Table  \ref{tab:Int_CD} below:

\begin{table}[h!]
\caption{EB estimates of intensive CD \label{tab:Int_CD}}
\centering
\begin{tabular}{|l|c|c|c|}
\hline
 & Chile & Colombia & Japan \\
\hline
Mean of a & 5.02 & 4.21 &  2.28 \\
\hline
Std of a & 1.22 & 1.16 &  1.25 \\
\hline
Mean of b & 0.40 & 0.39 & 0.38 \\
\hline
Std of b & 0.23 & 0.27 & 0.20 \\
\hline
Corr of a,b & -0.86 & -0.87 & -0.87 \\
\hline
\end{tabular}
\end{table}
which shows a large heterogeneity in the production parameters in all three samples, and again a consistently large negative correlation between the factor neutral parameter and the factor output elasticity.

%%%%%%%%%%%%%%%%%%%%%%%%%%%%%%%%%%%%%%%%%%%%%%%%%%%%%%%%%%%
   
\subsection{Unexplained heterogeneity}

We have consistently found a large  inter-firm heterogeneity in estimated technology parameters. This naturally raises the question: what proportion of this heterogeneity is linked to observable firm characteristics such as firm size or production sector?

The first row in Table \ref{tab_naceK} shows that in a simple variance decomposition (ANOVA with 33, 43, 39 degrees of freedom for the three countries), the share of the variance explained by size (10 groups for the deciles of Y, K, L, and respectively 8, 17, 13 industry 2-digit level industries for Chile, Colombia and Japan) is less than 10\% for the $\alpha, \beta, \gamma$ CD parameters, with the exception of intercept in Japan (around 13\%). The share of the variance explained by industry sectors alone is very small, with a peak of 2\% for the intercept in Colombia.
%The ANOVA decomposition for the CES case depicts a very similar picture. 

\vspace{0.3cm}

\begin{table}[hbt]
		%\vspace{}
		\centering
		\caption{Fraction of variance explained by size and industry \label{tab:pw_naceK}}
		\begin{adjustbox}{width=0.5\textwidth}
			\begin{threeparttable}
			%\small{
				\begin{tabular}{l|cccc}
					\toprule
					&\multicolumn	{4}	{c}{CHILE}	\\
					\hline
					&$\alpha^0$&$\beta$&$\gamma$ &\\
					\midrule
					\\
						Explained by sectors and size  &8.9\%  &6.4\%&6.0\%&\\
\\
					Explained by sectors  &0.5\%  &0.9\%&1.0\% &\\
					\\
					\hline
					&\multicolumn	{4}	{c}{COLOMBIA}	\\
					\hline
					&$\alpha^0$&$\beta$&$\gamma$ &\\
					\midrule
					\\
						Explained by sectors and size  &7.3\%  &7.2\%&5.3\%&\\
\\
					Explained by sectors  &2.1\%  &1.5\%&1.1\% &\\
					\\
					\hline
					&\multicolumn	{4}	{c}{JAPAN}	\\
					\hline
					&$\alpha^0$&$\beta$&$\gamma$ &\\
					\midrule
					\\
						Explained by sectors and size  &7.3\%  &4.7\%&5.2\%&\\
\\
					Explained by sectors  &1.9\%  &0.9\%&0.6\% &\\
					\\

		\bottomrule
				\end{tabular}\medskip{}

					\footnotesize
					\vspace{-3mm}
			\end{threeparttable}
		\end{adjustbox}
		\label{tab_naceK}
	\end{table}

We also look at the heterogeneity of estimated parameters at the industry sector levels. We see similar patterns both for the CD and CES estimated parameters (figures \ref{CD_nace} and \ref{CES_nace} in the Appendix).
The general message is that most inter-firm technological differences are within, not across, size or sectors. Attempting to fully address heterogeneity by assuming sector or size homogeneity appears to be rather unhelpful.

%%%%%%%%%%%%%%%%%%%%%%%%%%%%%%%%%%%%%%%%%%%%%%%%%%%
%%%%%%%%%%%%%%%%%%%%%%%%%%%%%%%%%%%%%%%%%%%%%%%%%%

\section{Implications of technological heterogeneity}\label{learn}

%%%%%%%%%%%%%%%%%%%%%%%%%%%%%%%%%%%%%%%%%%%%%%%%%

\subsection{The puzzle of negative correlation between factor neutral productivity and returns to scale}\label{ss:neg}

As we stressed in the previous sections, we found in all of our estimations (under CD, CES and Intensive CD production functions) a rather striking negative correlation between the intercept and returns to scale. Figure \ref{NegCorr} below shows the scatterplot in the CD case.
\begin{figure}[htp]\captionsetup[subfigure]{font=footnotesize}
\centering
%\begin{subfigure}[b]{.50\textwidth}
\centering
\includegraphics[width=.80\textwidth]{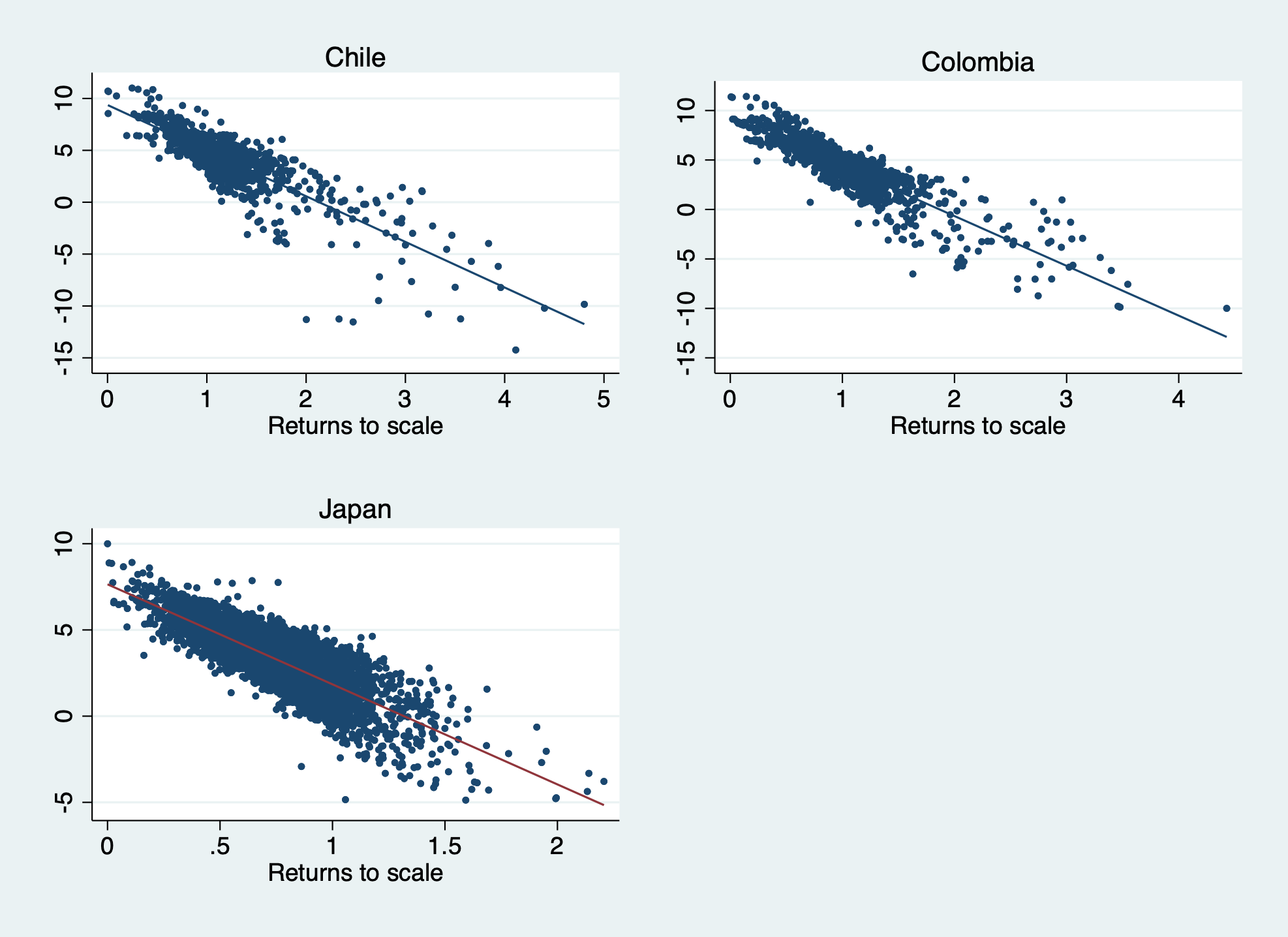}
\caption*{CD}\label{fig:CD_Corr}
\caption{Returns of scale vs Intercepts}\label{NegCorr}
\end{figure}

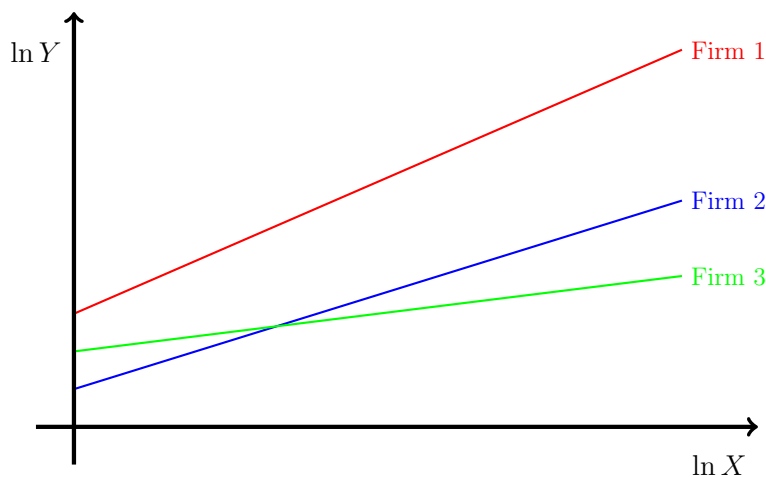
\begin{figure}[htp]
\centering
\begin{tikzpicture}
\draw[help lines, color=gray!30, dashed] (-4.8,-4.8) ;
\draw[->,ultra thick] (-4.5,2)--(5,2) ;
\draw [-,thick](4,1.5)node[right, scale=0.9]{$\ln X$};
%\draw[7.5,-1];
\draw[-, thick,color=red] (-4,3.5)--(4,7) node[right, scale=0.8]{Firm 1};
\draw[-, thick,color=blue] (-4,2.5)--(4,5) node[right, scale=0.8]{Firm 2};
\draw[-, thick,color=green] (-4,3)--(4,4)node[right, scale=0.8]{Firm 3} ;
\draw[->,ultra thick] (-4,1.5)--(-4,7.5) ;
\draw [-,thick](-4.5,6.7)node[above, scale=0.9]{$\ln Y$};
\end{tikzpicture}
\vspace{-6cm}
\caption{Dominated production functions}\label{Domin}	
\end{figure}

 We propose an explanation for this apparently puzzling result in terms of dominated production techniques. Consider a simple one-factor production function $\ln Y = a + b \ln X$. In Figure \ref{Domin} below  we plot the production function for three firms with heterogeneous parameters $(a_i,b_i)$ for $i=1,2,3$. Since firm 1 has both a higher intercept \textit{and} slope than firms 2 and 3, its production technology dominates those of the other two firms. Generally, given a set of firms $\{a_n,b_n\}_{n=1:N}$, absence of dominance implies the condition $(a_i-a_j)(b_i-b_j)\leq 0$ for all $i,j$ (i.e. a negative correlation between $a$ and $b$).
 
On the other hand, assuming heterogeneity in factor neutral productivity $a$ but homogeneity in factor output elasticity $b$ implies a strict order of dominance in production technologies for all firms in the market, which is not compatible with any reasonable concept of firms' equilibrium. On the contrary, absence of dominance implies that some firms employ  better factor neutral technologies, and others have better returns from the inputs.  This is reflected  in the large negative correlation between $a$ and $b$ found in our estimation of the intensive production functions in section \ref{intensive} above.

In the presence of more than one input, one possible way to address production dominance is to define for each firm $i=1\dots N$ a `composite input', say $\bar{X}_{it} = g(K_{it},L_{it})$ for an appropriate homogeneous function $g: \mathbb{R}^2_{+} \mapsto \mathbb{R}_{+}$, and then follow the same reasoning as in the single input case.
\citet*{hardy1952inequalities} famously discuss a family of homogeneous functions for aggregating sets of numbers called the \textit{generalized means of order $p$}, $(\sum_k a_k \theta_k^p)^{1/p}$, which encompasses most usual definitions of means. 

Defining the `composite input' $\bar X$ as a generalized mean of order $p$, we may rewrite both the CD and CES estimating equations \eqref{eq:final} and  \eqref{eq:est} as 
\[
\ln Y_{it} = \bar{\alpha}_i + \nu_i \ln (\bar{X}_{it})+ \upsilon_{it},
\]
with $\bar{X}_{it} = \left ( \omega_i K_{it}^\frac{\sigma_i-1}{\sigma_i}  + \left (1-\omega_i \right) 
L_{it}^\frac{\sigma_i-1}{\sigma_i}  \right )^{\frac{\sigma_i}{\sigma_i-1}}$
 in the CES case, and $\bar{X}_{it}  = K_{it}^{\frac{\beta_i}{\nu_i}} L_{it}^{\frac{\gamma_i}{\nu_i}}$ in the CD case (the limiting case of the generalized mean of order $p$ for $p \to 0$). The non dominance condition  $(\bar{\alpha}_i-\bar{\alpha}_j)(\nu_i-\nu_j)\leq 0$ induces a negative correlation between the factor neutral parameter 
 $\bar{\alpha}$ and the returns to parameter scale $\nu$, which we consistently find in both our CD and CES estimations and in  all three samples.

We notice the link to the theoretical literature on technological menu choice (\citealp{jones2005shape}, \citealp*{leon2019appropriate}). In this literature firms choose an appropriate technology 
(the local production function) from an available technological bundle (the global production function) in the 
technology parameter space. When the isoquant in this space has a well behaved concave shape, if firms do not choose dominated technologies, a negative correlation would appear.%
 \footnote{As argued above, the standard strategy of allowing heterogeneity in the intercept but assuming homogeneity in the output elasticities implies a strict order of dominance of local production functions which appears to be in contrast with the technology menu literature.}

%%%%%%%%%%%%%%%%%%%%%%%%%%%%%%%%%%%%%%%%%%%%%%%%%%%%%%%%%%%%%%%%%%%%%%%%%%%%%%%%%%

\subsection{Productivity measurement}

In a well-known survey on productivity measurement,  \citet{syverson2011determines} describes large and persistent measured productivity differences across firms as a stylized fact. In the standard estimating equation $y_{it} = a_i + b k_{it} + c l_{it} + \epsilon_{it}$, $a_i$ is used as the (logged) productivity estimate for firm $i$, typically referred by a variety of names such as ``Total Factor'', ``Factor Neutral'', or ``Hicksian'' productivity. Reviewing studies over the past couple of decades, \citet{syverson2011determines} argues that  ``\textit{researchers in many fields [....] have documented, virtually without exception, enormous and persistent measured productivity differences across producers, even within narrowly defined industries}''.
  
 In our three datasets a measure of the  dispersion of factor neutral productivity may be obtained, following \citet{syverson2011determines}, by comparing the values of $e^{a}$ at the 90th and 10th percentile. Using the values of the estimated $\hat{a}_i$ from our estimation, using the two digits industry level we find average 90/10 differences within each sector equal to 188.8, 94.3 and 53.7 in Chile, Columbia and Japan. respectively
 %\footnote{DA CHIARIRE QUESTA FOOTNOTE, SIAMO SIVURI LA VOGLIAMO LASCIARE? COSI' NON MI E' MOLTO CHIARA  \citet{hsieh2009misallocation}, for instance, show that Indian and Chinese firms have much more dispersed TFP distributions than U.S. firms. Country and sector effects may introduce heterogeneity through mechanisms such as varying factor constraints (e.g., credit markets), competition levels, and proximity to the technology frontier. CHE VUOL DIRE?  Mixing heterogeneous countries would result in heterogeneous parameters, even in the absence of within-country heterogeneity. A similar argument applies to sector heterogeneity or long panels.}. 
 
 Even taking  into account the fact that at two digit industry  level we may expect significant differences in firms TFPs, these numbers are indeed enormous. As argued above, if we were appraising firms' productivity simply by using  
 $\hat{a}_i$ without taking into account factors' output elasticities heterogeneity, we would implicitly assume 
 a strict order of dominance in firms' adopted technologies. Clearly these numbers are hardly compatible with any reasonable concept of industry equilibrium.
 
The standard practice in the literature of using factor neutral productivity to appraise 
firms' heterogeneity in technological productivity does not account for heterogeneity in the factors' elasticities of output.  A proposal for considering {\textit both} factors output elasticities and  factor neutral productivity for productivity comparisons is discussed in \citet{bernard1996comparing}, who propose a measure of productivity called Total Technology Productivity (TTP). Given the estimating equation $y_{it} = \alpha_i + \beta_i k_{it} + \gamma_i l_{it} + \epsilon_{it}$, we  follow
\citet{bernard1996comparing} and define %
 \[
 \ln (TTP_i) =  \alpha_i + \beta_i \ln (K_0) + \gamma_i \ln (L_0)
 \]
 for an appropriate choice of reference levels of capital and labor $K_0$ and $L_0$. In words, TTP captures how much any given firm would produce if all firms employed exactly the same amount of $K_0$ and $L_0$ of capital and labor.  We calculate $\ln (TTP)$ for each sector, using the median values in each sector of capital and labor for $K_0$ and $L_0$, as recommended by \citet{bernard1996comparing}. The results are illustrated in figure \ref{TTP} below, where the yellow histograms refer to the distribution of (logged) factor neutral productivity, the blue ones the distribution of returns to scale, and the red ones $\ln (TTP)$.

  Looking at the histograms, it appears that both (logged) factor neutral productivity and returns to scale  have  roughly comparable dispersion.

\begin{figure}[H]

\centering
%\vspace{-2cm}
\begin{minipage}[b]{.49\textwidth}
\centering
\includegraphics[width=.99\textwidth]{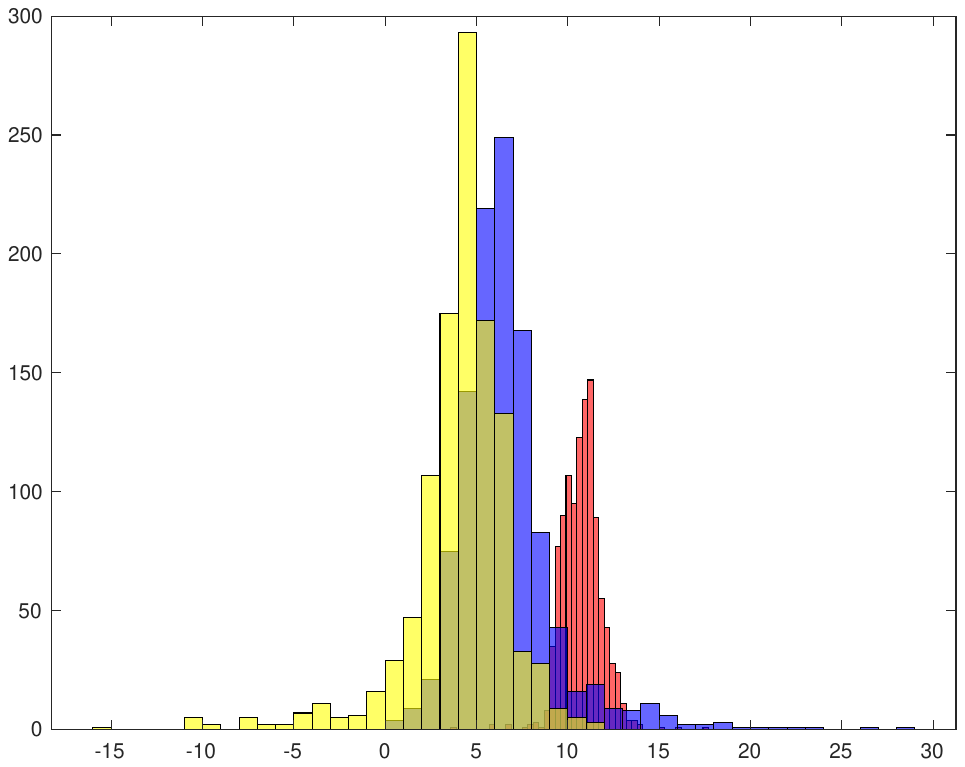}
\vspace{-3.5cm}
\caption*{Chile}\label{fig:Chile by sectors}
\end{minipage}
\begin{minipage}[b]{.49\textwidth}
\centering
\includegraphics[width=.99\textwidth]{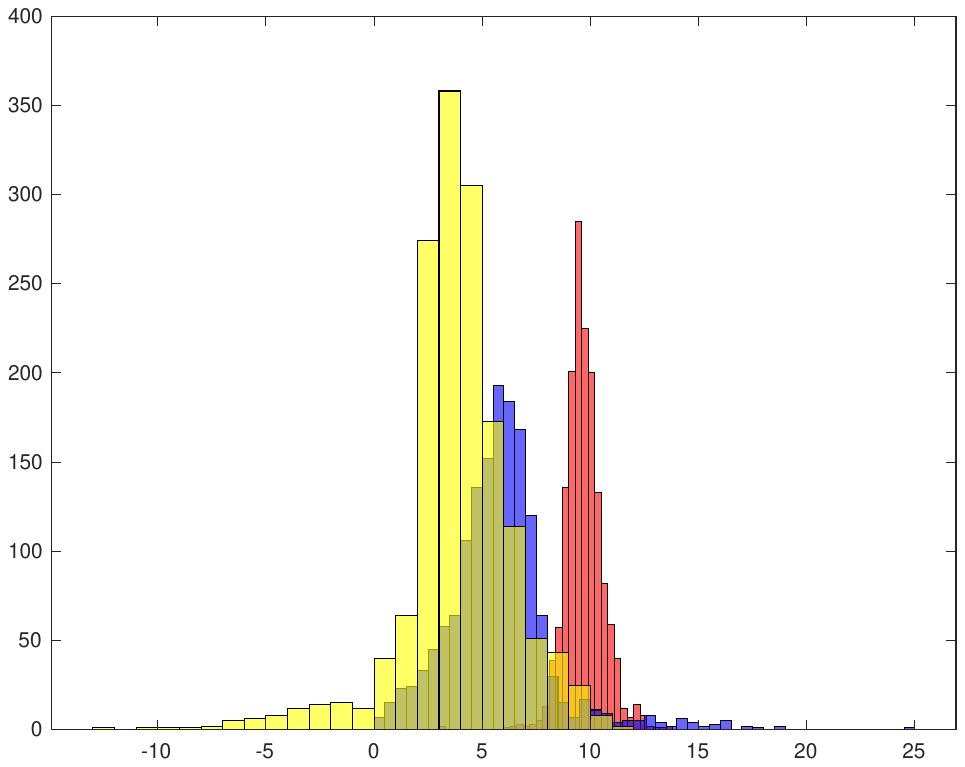}
\vspace{-3.5cm}
\caption*{Colombia }\label{fig:Colombia by sectors}
\end{minipage}
\begin{minipage}[b]{.49\textwidth}
\centering
\vspace{-2cm}
\includegraphics[width=.99\textwidth]{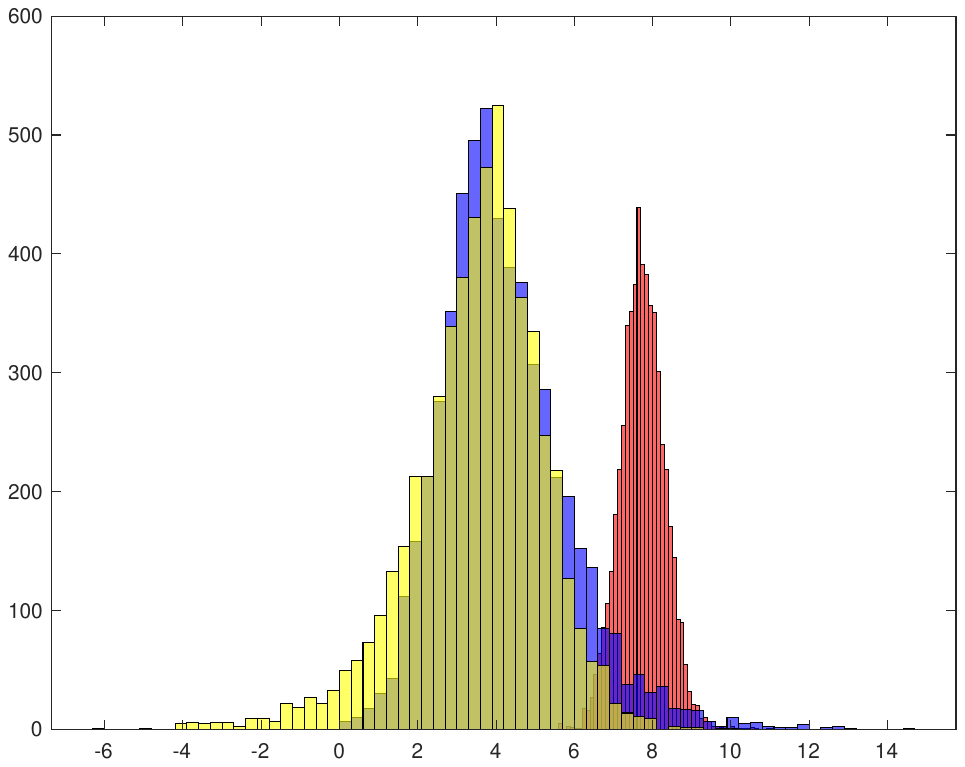}
\vspace{-3.5cm}
\caption*{Japan}\label{fig:Japan by sectors}
\end{minipage}

\caption{TTP distributions}\label{TTP}

\end{figure}

The mean of $\ln (TTP)$ equals the sum of the means of the two types of productivity;  however, given the strong negative correlation between $\alpha$ and $\beta + \gamma$,  the histograms of $\ln (TTP)$ show a dramatic reduction
in total productivity dispersion: the 90/10 ratios  of $\ln (TTP)$ are equal to 3.92, 3.16 and 3.88 respectively for Chile,  Columbia and Japan, which, even at the rather coarse sector level, seem much more reasonable estimates of productivity dispersion than those calculated without considering factors' productivity.\footnote{As for intercountry differences, notice that there are 7,  17 and 13 two-digit sectors in Chile, Columbia and Japan.  The standard deviation of $\ln (TTP)$ is respectively 38\%, 35\% and 34\% smaller than the logged factor neutral productivity $\ln (\alpha)$. For robustness, we repeat these productivity calculations using the CES production function, and in the Appendix we show that results are comparable to those obtained using the Cobb Douglas.}

Finally, notice that the  $\ln(TTP)$ decomposition is not directly applicable to the Translog production function, since heterogeneity in individual firms output elasticities is induced by different factors' levels, so that for any fixed level of $K_0$ and $L_0$ there is no production heterogeneity. However, it is interesting to notice, from Table \ref{tab:pw} above, that the 90/10 ratio in TFP ($\exp  \alpha_i$) is equal to  3.94, 3.60 and 3.01  in Chile,  Columbia and Japan, which are quite comparable to our estimates of the 90/10 ratios for TTP. Of course, the key difference is that Translog productivity dispersion estimates attribute these productivity differences solely to the factor neutral components. 
 
%%%%%%%%%%%%%%%%%%%%%%%%%%%%%%%%%%%%%%%%%%%%%%%%%%%%%%%%%%%%%%%%%%%%%%%%%%%%%%%%%%

\subsection{Markup Estimation}

In the production approach, the markup for any input is given by the ratio of its output elasticity to its share of revenue (\citealp{hall1988relation}, \citealp{de2011product}). 
Different ways to account for technological parameter heterogeneity may result in different markups: \citet{raval2023testing} observes that, "\textit{...in order to use the production approach, economists will have to allow more heterogeneity in technology}" as ignoring such heterogeneity can result in systematically biased markups estimation. For instance, under a homogeneous CD production function, all firms have the same output elasticities, so differences in markups will only arise from differences in inputs' share of output. Conversely, as discussed in the previous subsection \ref{ss_CD_het}, the Translog production function underestimates the extent of heterogeneity, which will affect markups estimation.

\begin{figure}[H]
\centering
\includegraphics[width=.90\textwidth]{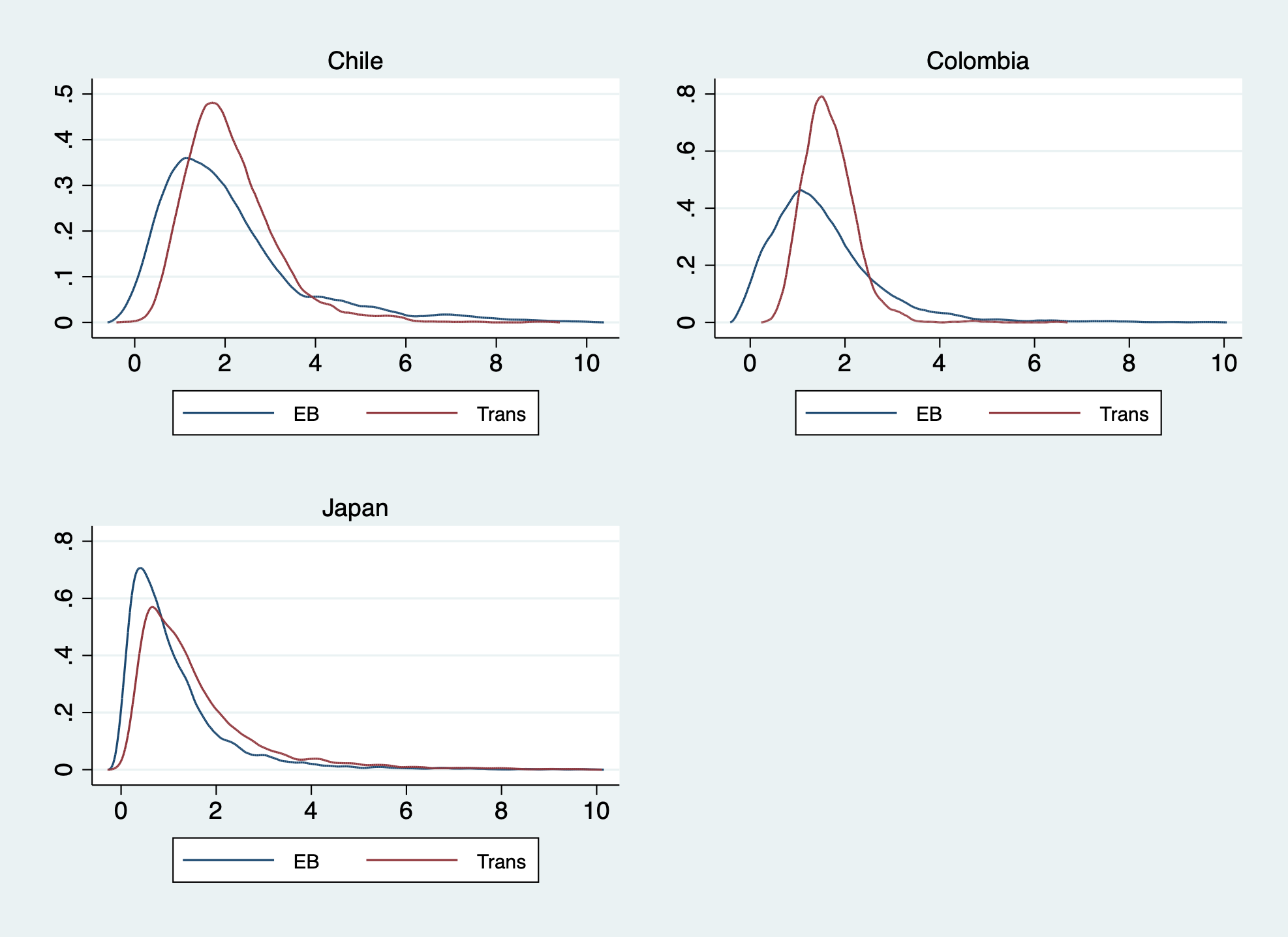}
\caption{Markups comparison}\label{Mkup_fig}
\end{figure}

We compute labor markups using under two methods: i) using our EB estimated firm specific labor elasticity coefficients, and ii) using the estimated  firm specific labor elasticity coefficients from the ACF Translog. Results are presented in Figure \ref{Mkup_fig} and Table \ref{tab:mark_ups}. The EB measures are much more dispersed (for Chile and Colombia the standard deviation and the 90/10 percentile ratios are both around double in EB case and slightly less for Japan) due to the detection of higher levels of heterogeneity. 
%XXX QUESTO LO LEVEREI PERCHE' PER GIAPPONE NON E' VERO: 
%This is an further confirmation that our EB procedure in general is useful in detecting parameters dispersion, while at the same time providing measures of central tendency consistent with other approaches.

\begin{table}[h]
\centering
     %\begin{adjustbox}[width=0.7\textwidth]
        \begin{tabular}{llccccc}
            \hline
             \multicolumn{2}{l}{}&Mean&St.Dev.&90/50 perc.&90/10 perc.&Firms\\
                          \hline
            \multirow{2}{*}{Chile} & Translog &\multicolumn{1}{c}{2.13} & \multicolumn{1}{c}{1.00}& \multicolumn{1}{c}{1.71}& \multicolumn{1}{c}{3.16}& \multicolumn{1}{c}{1207} \\
                                 &EB& \multicolumn{1}{c}{2.18} & \multicolumn{1}{c}{1.88}& \multicolumn{1}{c}{2.52}& \multicolumn{1}{c}{7.64}& \multicolumn{1}{c}{1096} \\
                                 \hline
            \multirow{2}{*}{Colombia} & Translog &\multicolumn{1}{c}{1.70} & \multicolumn{1}{c}{1.37}& \multicolumn{1}{c}{1.44}& \multicolumn{1}{c}{2.27}& \multicolumn{1}{c}{1760} \\
                                 &EB& \multicolumn{1}{c}{1.64} &\multicolumn{1}{c}{2.42}& \multicolumn{1}{c}{2.24}&  \multicolumn{1}{c}{8.31}&\multicolumn{1}{c}{1535} \\
                                 \hline
            \multirow{2}{*}{Japan} & Translog &\multicolumn{1}{c}{1.72} & \multicolumn{1}{c}{2.41}& \multicolumn{1}{c}{2.66}& \multicolumn{1}{c}{6.89}& \multicolumn{1}{c}{5588} \\
                                 &EB& \multicolumn{1}{c}{1.25} & \multicolumn{1}{c}{2.50}& \multicolumn{1}{c}{2.96}& \multicolumn{1}{c}{10.95}& \multicolumn{1}{c}{5588} \\
                                 \hline        
                                 \end{tabular}                  
    % \end{adjustbox}
   \caption{Labor markups for Chile, Colombia and Japan}
    \label{tab:mark_ups}
\end{table}

\section{Conclusions}\label{conclusion}

%The paper shows that the Empirical Bayes approach effectively addresses heterogeneity in estimating production functions, revealing important characteristics of the production technology.
%Using a nonparametric EB estimation method, we uncovered significant inter-firm heterogeneity in both factor-neutral productivity and output elasticities of factors, and a strong negative correlation between these two components. These results are consistent across the three samples of Chilean, Colombian and Japanese firms, and are robust to using CD, CES and intensive CD production functions.

%We show that taking into account inter-firm technological heterogeneity is very useful to address empirical issues such as productivity heterogeneity, misallocation and markups measurement. It also opens the way to more direct and empirically founded links with related theoretical issues such as local technology appropriateness (proposed at least since \citealp{atkinson1969new}) and technology menu (\citealp{jones2005shape}, \citealp{leon2019appropriate}).

 %It has to be noted that our EB method has certain limitations. 
The paper shows that the Empirical Bayes (EB) approach effectively addresses heterogeneity in estimating production functions, thereby revealing key characteristics of production technology. By employing a nonparametric EB estimation method, we identified substantial inter-firm heterogeneity in factor-neutral productivity and output elasticities of inputs, as well as a large negative correlation between these two aspects. These findings are consistent across datasets from Chilean, Colombian, and Japanese firms and remain robust when applying CD, CES, and intensive CD production function specifications.

We show that taking into account inter-firm technological heterogeneity is  beneficial for addressing empirical issues such as productivity, misallocation and markup measurements. It also enables a more direct and empirically grounded connection with related theoretical concepts such as local technology suitability (discusses at least since Atkinson and Stiglitz, 1969) and technology menu choice (Jones, 2005; León-Ledesma and Satchi, 2019).

However, it is important to note some limitations of our EB method. A key requirement for identifiability is that each firm must have more observations than the number of parameters being estimated. This implies that more complex models require longer panel data. Nonetheless, even with extended panels, the curse of dimensionality quickly constrains the feasible size of the discrete grid.

\singlespacing 
\subsubsection*{Acknowledgements.} 
We thank the participants of the Productivity and Growth Workshop in Palermo, and in particular to Daniel Ackerberg, Mert Demirer, Miguel Leon-Ledesma and Devesh Raval for valuable discussions and suggestions.  We are also grateful to Christopher Parmeter, Nicola Persico, David Rivers, and Chad Syverson for their comments, and to Cinzia Di Novi for kind hospitality.
 \newpage
\bibliography{EB.bib}

\clearpage

%%%%%%%%%%%%%%%%%%%%%%%%%%%%%%%%%%%%%%%%%%%%%%%%%%%%%%%%%
\appendix

\section{Heterogeneous CES production function}\label{Het_CES}

The CD production function implicitly assumes that the elasticity of substitution between capital and labor is unitary. 
For completeness, and to show the general applicability of our EB approach, we also estimate a well known generalization of the CD,  the CES  production function proposed by \citet{arrow1961capital}, hereon ACMS.

Briefly after ACMS, several authors proposed a generalization of the CES function accounting for the degree of homogeneity. Taking logs and assuming a heterogeneous dynamic structure for intercepts as in equation \eqref{eq:6}, with individual terms $\alpha^0, \alpha^1, \alpha^2$ the estimating equation is
\begin{equation}\label{eq:est}
y_{it} = \alpha^0_i + \alpha^1_{i}t + \alpha^2_{i}t^2 + \nu_i \left ( \frac{\sigma_i}{\sigma_i-1} \right ) \ln \left ( \omega_i K_{it}^\frac{\sigma_i-1}{\sigma_i}  + \left (1-\omega_i \right) L_{it}^\frac{\sigma_i-1}{\sigma_i}  \right ) + \upsilon_{it}
\end{equation}
where $\sigma$ is the elasticity of substitution between capital and labor,
$\omega$ is a distribution parameter, $\nu$  is the return to scale parameter and 
with $\upsilon_{it} \sim N(0,s_i)$.  Similarly to the CD case, the firm specific factor-neutral productivity
 $\bar{\alpha}_i=\frac{1}{T}\sum^T_{t=1} \alpha^0_i + \alpha^1_{i}t + \alpha^2_{i}t^2$.
 
This function, which is usually called the generalized CES production function, is homogeneous of degree $\nu$. \citet{maddala1967estimation} showed that, using the \citet{kmenta1967estimation} approximation, omitting $\nu$ may imply a substantial estimation bias in $\sigma$.\footnote{The original ACMS  formulation implicitly assumes constant returns to scale, and is the most popular version in applications. An alternative generalization focuses on introducing factor augmenting technological parameters, without the unitary sum constraint (introduced by \citealp{david1965biased}). We also estimated this second specification, obtaining quite similar results for the common parameters.}
This CES function has seven parameters ($\alpha^0,  \sigma, \nu, \omega, \alpha^1, \alpha^2, s$). We discretize $\alpha^0$ with 9 equally spaced points in $[-5,10]$, $\sigma$ with 9 points, $\nu$ with 9 points and $\omega$ with 9 points. For $s$ we use 6 points and we use a setting equal to the CD case for the $\alpha^1, \alpha^2$ terms. This results in $9^4*6^3 = 1,417,176$ parameters' configurations which nonparametrically approximate the true joint parameter distribution. We combine equations \eqref{eq:density} and \eqref{eq:est} to find the matrix $F$ which collects the conditional densities for all firms, and apply an iterative algorithm (starting with an uniform distribution)  to find the fixed point $\pi^*$. Table \ref{tab_CES} below reports the means of the estimated coefficients.
\vspace{0.5cm}

\begin{table}[hbt]
		\vspace{-2mm}
		\centering
		 \caption{Average value of CES parameters\label{tab:pw}}
		\begin{adjustbox}{width=0.7\textwidth}
			\begin{threeparttable}
				\begin{tabular}{lccccccc}
					\toprule
					& $\alpha^0$ & $\omega$& $\nu$  &$\sigma$&$\alpha^1$&$\alpha^1$&$s$\\
					\midrule
					\multirow { 2}{*}{Chile} & 4.590  & 0.360 & 1.166&2.605 & 0.041 & -0.002 & 392\\
					& (0.085) & (0.007) & (0.016) & (0.078) & (0.005) & (0.000)&(0.005)\\

					\hline
					Observations &  \multicolumn{7}{c}{12056}\\
					Firms&  \multicolumn{7}{c}{1096}\\
					\hline
										\multirow { 2}{*}{Colombia} & 4.562  & 0.200 & 1.047 & 2.759 & -0.058 & 0.005&0.370 \\
					& (0.053) & (0.004) & (0.009) & (0.051) & (0.003) & (0.000)&(0.004)\\
					%\\
					\hline
					Observations &  \multicolumn{7}{c}{18420}\\
					Firms&  \multicolumn{7}{c}{1535}\\
					\hline
										\multirow { 2}{*}{Japan} & 4.013  & 0.342 & 0.691 & 1.736 & 0.013&-0.0001 & 0.192\\
					& (0.017) & (0.002) & (0.003) & (0.008) & (0.001) & (0.000)&(0.001)\\
					%\\
					\hline
					Observations &  \multicolumn{7}{c}{39116}\\
					Firms&  \multicolumn{7}{c}{5588}\\
					\bottomrule
				\end{tabular}\medskip{}
				\begin{tablenotes}
					%\singlespacing
					%\vspace{-5mm}
					\item \textit{Notes}: Standard errors in parentheses.
					{\footnotesize\par}
				\end{tablenotes}
					\footnotesize
					\vspace{-3mm}
			\end{threeparttable}
		\end{adjustbox}
		\label{tab_CES}
	\end{table}

\vspace{0.3cm}

The table shows that the mean elasticity of substitution $\sigma$ is much bigger than one for all the countries; the mean of the distribution parameter $\omega$ is close to one third, unless than for Colombia, and the mean of the homogeneity parameters $\nu$  are close to the CD estimates.
The dynamic parameters $\alpha^1,\alpha^2$ suggest again an average U-shape distribution for the intercept.

\begin{figure}[H]
\centering
  \includegraphics[width=.83\textwidth]{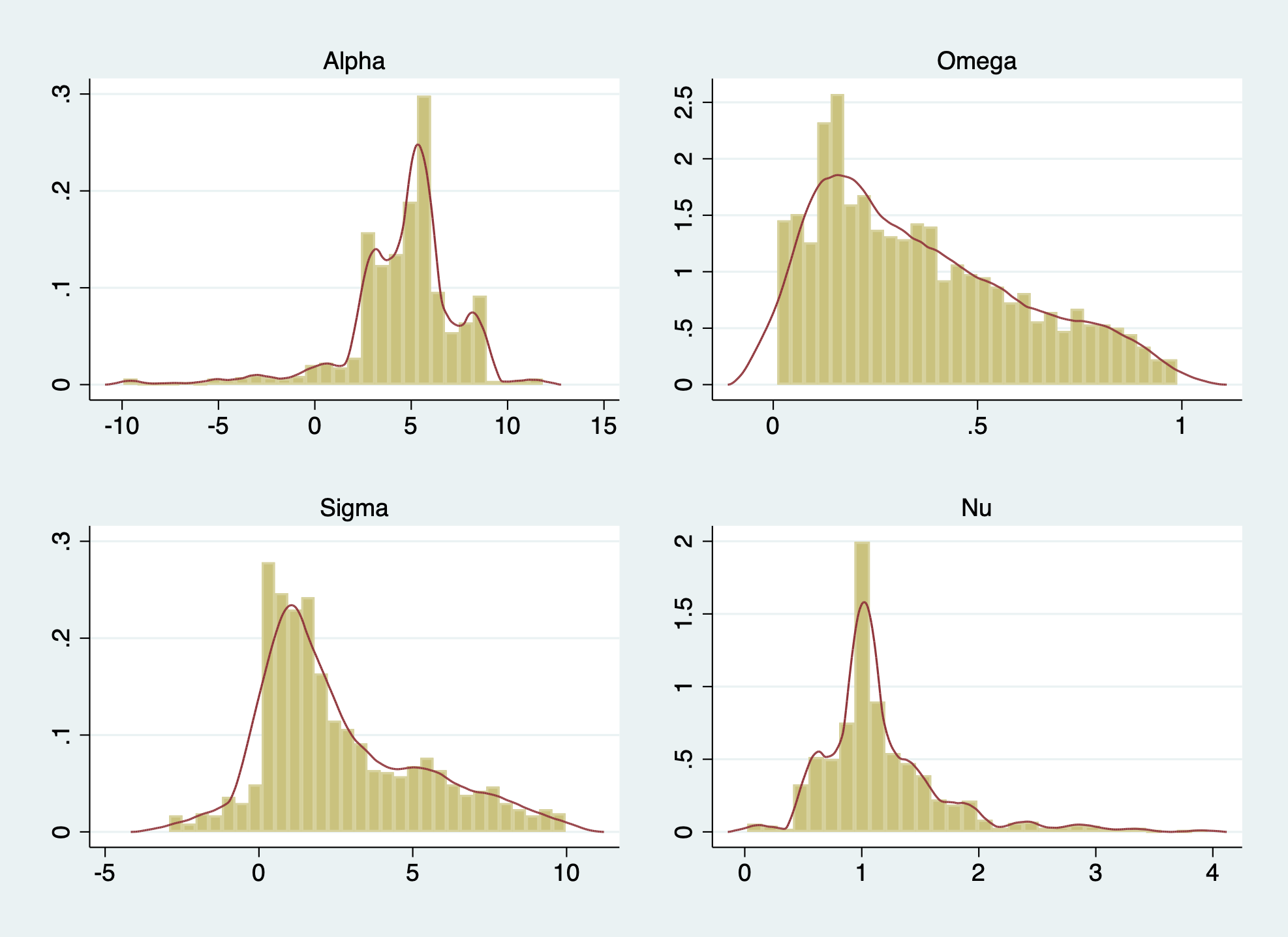}
  \caption{CES coefficients distributions Chile}\label{CHL_CES}
\end{figure}

\begin{figure}[H]
\centering
  \includegraphics[width=.83\textwidth]{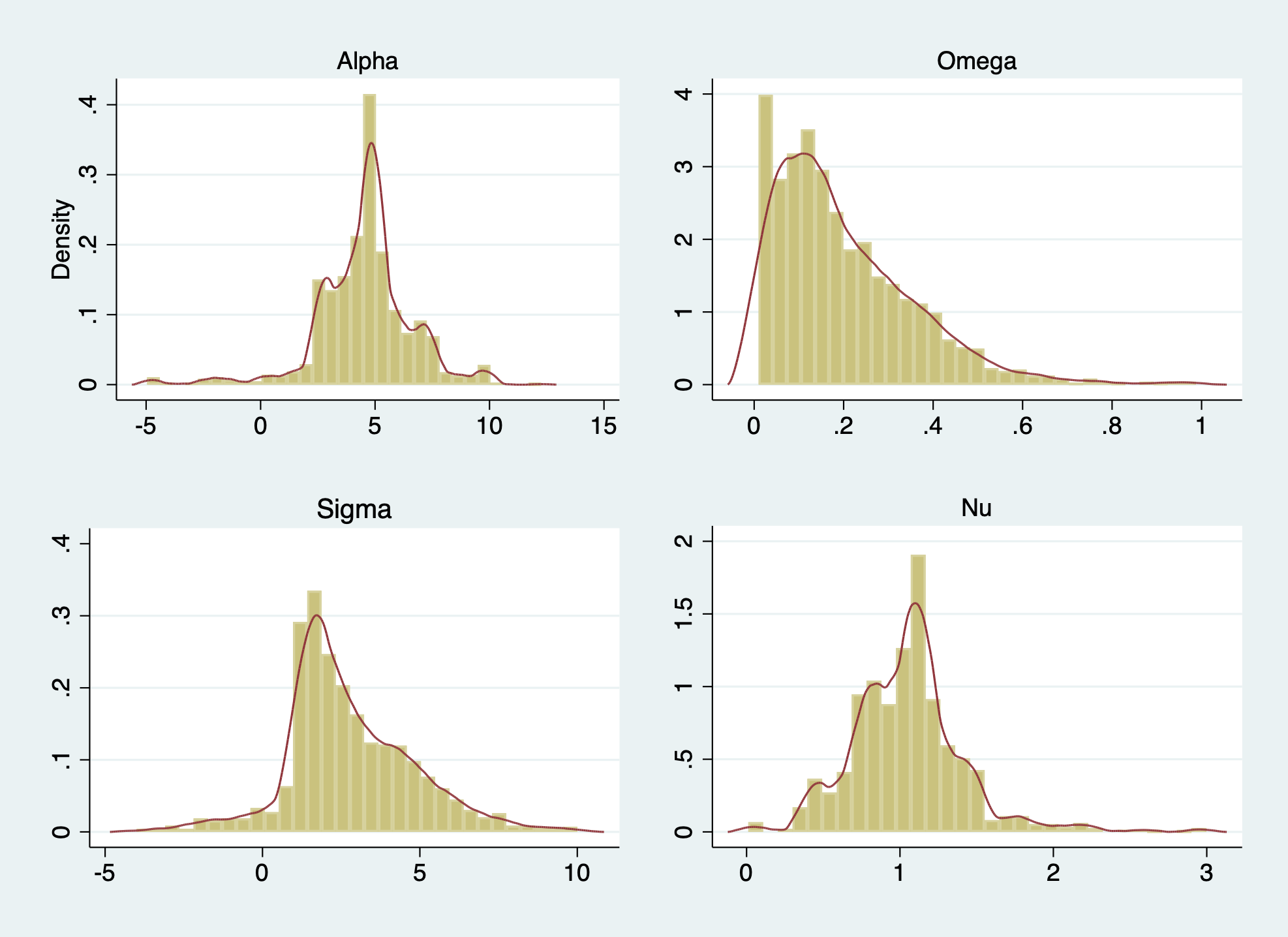}
  \caption{CES coefficients distributions Colombia}\label{COL_CES}
\end{figure}

\begin{figure}[H]
\centering
  \includegraphics[width=.83\textwidth]{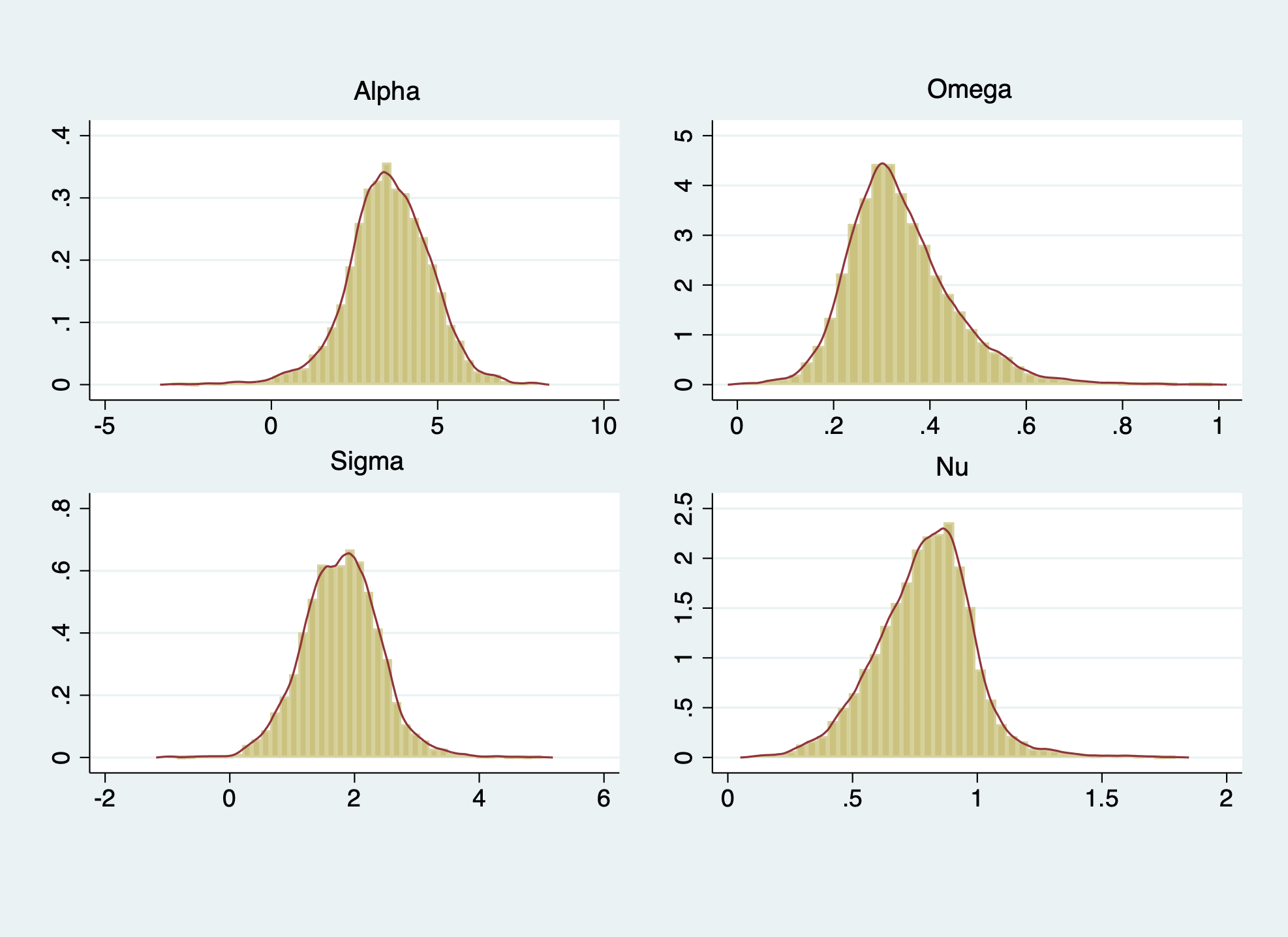}
  \caption{CES coefficients distributions Japan}\label{JAP_CES}
\end{figure}

From $\pi^*$ we get the posterior types joint distribution of the firm specific parameters. In Table \ref{tab_CES_Cor} we show the correlations of the parameters of major interest, and in Figures \ref{CHL_CES},  \ref{COL_CES},  \ref{JAP_CES} we present their histograms and (sorted) individual estimates which show again large heterogeneity in all technology parameters, similarly to the CD case. A large mass of the distributions of the elasticities of substitution are between 0 and 4, so that capital and labor may be seen more as substitutes than complements (but notice there are small proportion of firms in each countries that have complementary factors). 	As in the CD case, notice the very strong negative correlation between the intercept and  returns to scale.

%\vspace{.25cm}

	%\vspace{0.1cm}

	\vspace{1cm}
	
	\begin{table}[hbt]
		\vspace{-2mm}
		\centering

		\caption{CES: firm specific parameters correlation \label{tab:pw_distribution2}}
		\begin{adjustbox}{width=0.3\textwidth}
			\begin{threeparttable}

				\begin{tabular}{lcccc}
					\toprule

					\multicolumn{5}{c}{Chile}\\
					\hline
					&&$\omega$ &$\nu$&$\sigma$\\
					\hline
					&$\alpha^0$  &\multicolumn{1}{|c}{-0.195}&-0.822&-0.024\\
&$\omega$&\multicolumn{1}{|c}{}   & 0.071&-0.202\\
&$\nu$&\multicolumn{1}{|c}  {}    &&-0.106\\
\hline
\multicolumn{5}{c}{Colombia}\\
					\hline
						&&$\omega$ &$\nu$&$\sigma$\\
					\hline
					&$\alpha^0$  &\multicolumn{1}{|c}{-0.068}&-0.866&0.006\\
&$\omega$&\multicolumn{1}{|c}{}   & -0.245&-0.281\\
&$\nu$&\multicolumn{1}{|c}  {}    &&0.077\\
\hline

\multicolumn{5}{c}{Japan}\\
					\hline
					&&$\omega$ &$\nu$&$\sigma$\\
					\hline
					&$\alpha^0$  &\multicolumn{1}{|c}{0.006}&-0.748&-0.107\\
&$\omega$&\multicolumn{1}{|c}{}   & -0.145&-0.404\\
&$\nu$&\multicolumn{1}{|c}  {}    &&-0.145\\

\hline

					\bottomrule
				\end{tabular}\medskip{}

					\footnotesize
					\vspace{-3mm}
			\end{threeparttable}
		\end{adjustbox}
		\label{tab_CES_Cor}
	\end{table}

\begin{figure}[htp]
\centering
\includegraphics[width=.60\textwidth]{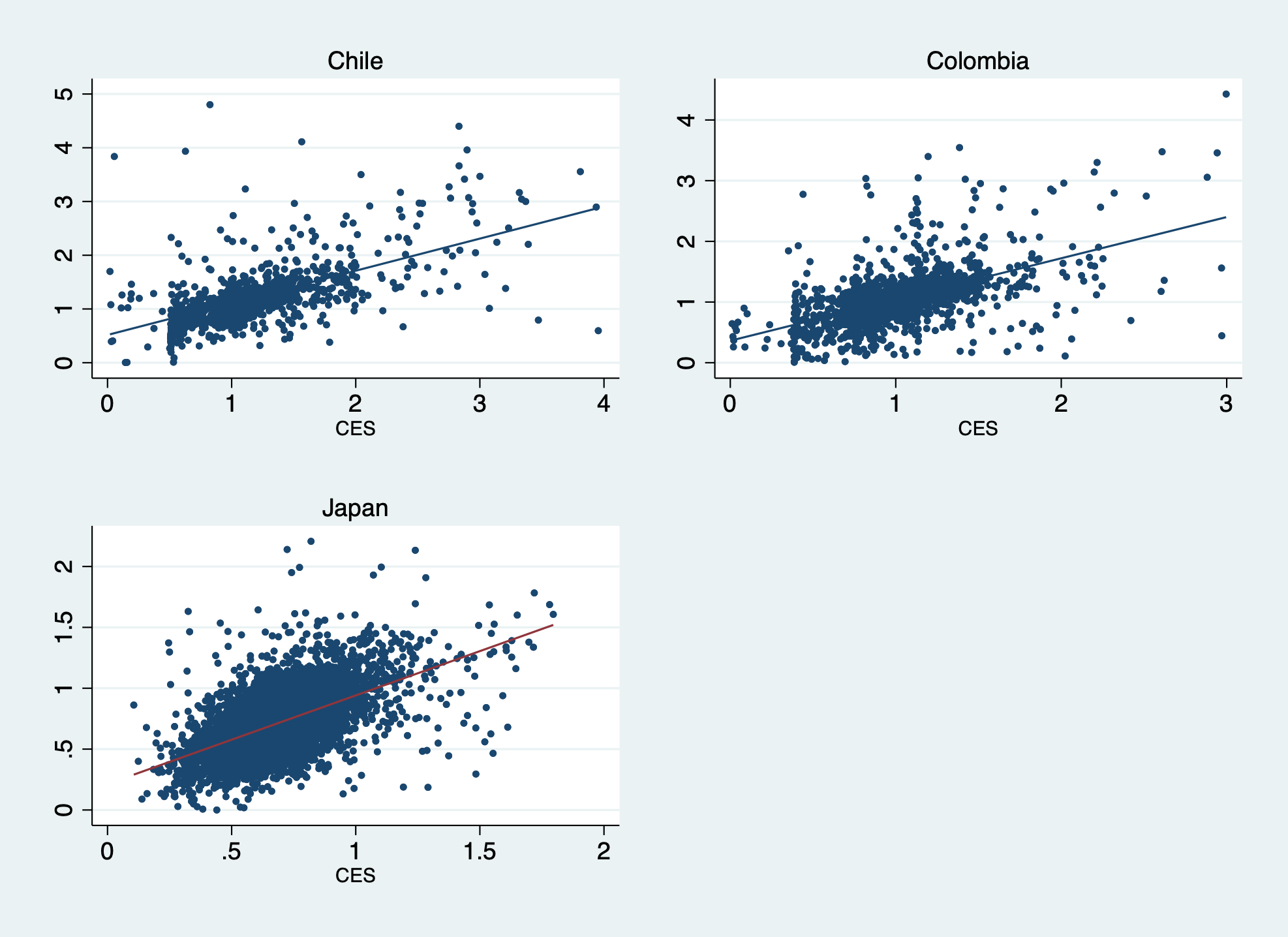}
\caption*{Returns to Scale}\label{fig:CD_CES_nu}
%\end{subfigure}%
%\begin{subfigure}[b]{.50\textwidth}
%\centering
%\includegraphics[width=.80\textwidth]{Corr_CES.png}
%\caption*{CES}\label{fig:CES_Corr}
%\end{subfigure}

\caption{CD-CES comparison}\label{CD_CES}

\end{figure}

\newpage

\section{Other results}

\begin{table}[hbt]
		%\vspace{-8mm}
		\centering
		%\resizebox{\columnwidth}{!}{
		\caption{Comparison among estimates' elasticities \label{tab:hom_est}}
		\begin{adjustbox}{width=0.5\textwidth}
			\begin{threeparttable}
			%\small{
				\begin{tabular}{l|l|ccc}
					\toprule
					%&\multicolumn{3}{c}{Full sample}&\multicolumn{3}{c}{Working sample}\\
					&& \textit{EB} & \textit{ACF Translog}&\textit{Pooled} \\
					\midrule

					%\\
					\multirow {5}{*}{Chile}& Capital-Output elasticity  &0.425 & 0.336  &0.476 \\
					
					&&(0.009) & (0.003) & (0.019) \\
					
			                &Labor-Output elasticity   &0.789& 0.816  & 0.713 \\
					
					&&(0.016) & (0.007) & (0.033) \\
					%\\
					\hline
					&RTS   &1.214   &1.149 & 1.189  \\
					%\\
%					$3^{rd}$ quartile   & 8.67  & 9.44&4.61&9.08&9.80&4.91 \\
%					%\\
%					Standard deviation  & 1.76 &1.86 &1.41 &1.50&1.66 &1.32 \\
%					%\\
				
				%	&Obs&12056&12056&12056\\
				%&Firms&1096&1096&1096\\
					\hline
									\multirow {5}{*}{Colombia}&Capital-Output elasticity   &0.407 & 0.262  & 0.473\\
					
					&& (0.007)& (0.002) & (0.015) \\
					
			                &Labor-Output elasticity   &0.668&0.728   & 0.718  \\
					
				&& (0.011)& (0.003) &(0.022) \\

					%\\
						\hline
					&RTS   &1.074&0.990 & 1.191 \\

					%\\
%					$3^{rd}$ quartile   & 8.67  & 9.44&4.61&9.08&9.80&4.91 \\
%					%\\
%					Standard deviation  & 1.76 &1.86 &1.41 &1.50&1.66 &1.32 \\
%					%\\
				
			%		&Obs&18420&18420&18420\\
			%	&Firms&1535&1535&1535\\
\hline
					\multirow {5}{*}{Japan}&Capital-Output elasticity   &0.329 &0.237&0.264\\
					
					&&(0.003) & (0.002) & (0.010) \\
					
			                &Labor-Output elasticity   &0.387& 0.543  & 0.600 \\
					
					&&(0.003) & (0.003) & (0.014) \\
					%\\
						\hline
					&RTS   &0.720   &0.780 & 0.864  \\
					%\\
%					$3^{rd}$ quartile   & 8.67  & 9.44&4.61&9.08&9.80&4.91 \\
%					%\\
%					Standard deviation  & 1.76 &1.86 &1.41 &1.50&1.66 &1.32 \\
%					%\\
					%\hline
			%		&Obs&39116&39116&39116\\
			%	&Firms&5588&5588&5588\\

					%\hline\\
					\bottomrule
					\end{tabular}\medskip{}
					\begin{tablenotes}
					\item Standard errors among parenthesis. Pooled contains year dummies. 
										%{\footnotesize\par}
				\end{tablenotes}

					\footnotesize
					\vspace{-3mm}
			\end{threeparttable}
		\end{adjustbox}
		\label{tab_stat}
	\end{table}

\begin{figure}[H]

\centering
%\vspace{-2cm}
\begin{minipage}[b]{.49\textwidth}
\centering
\includegraphics[width=.99\textwidth]{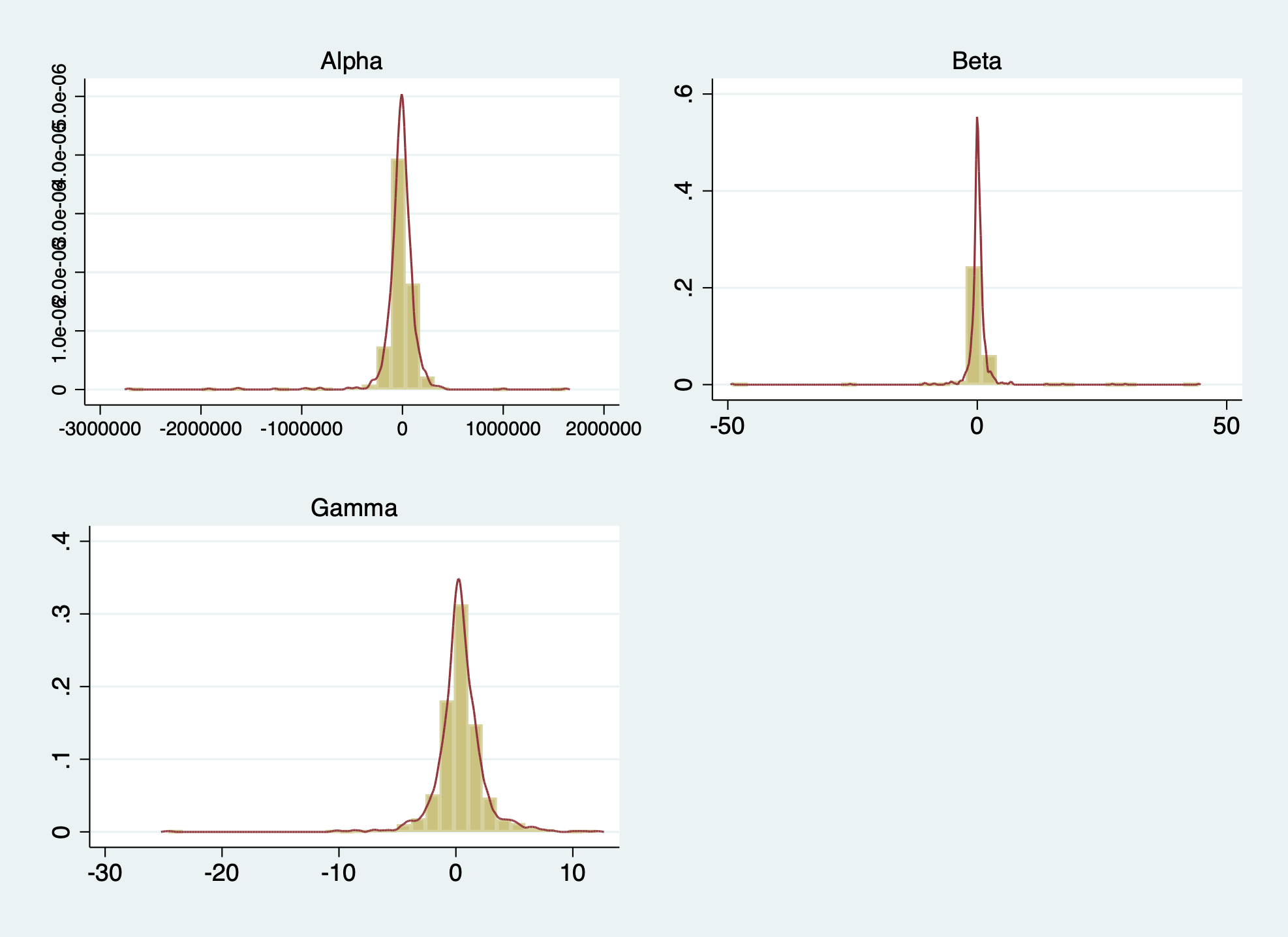}
%\vspace{-3.5cm}
\caption*{Chile}\label{fig:ChileOLS}
\end{minipage}
\begin{minipage}[b]{.49\textwidth}
\centering
\includegraphics[width=.99\textwidth]{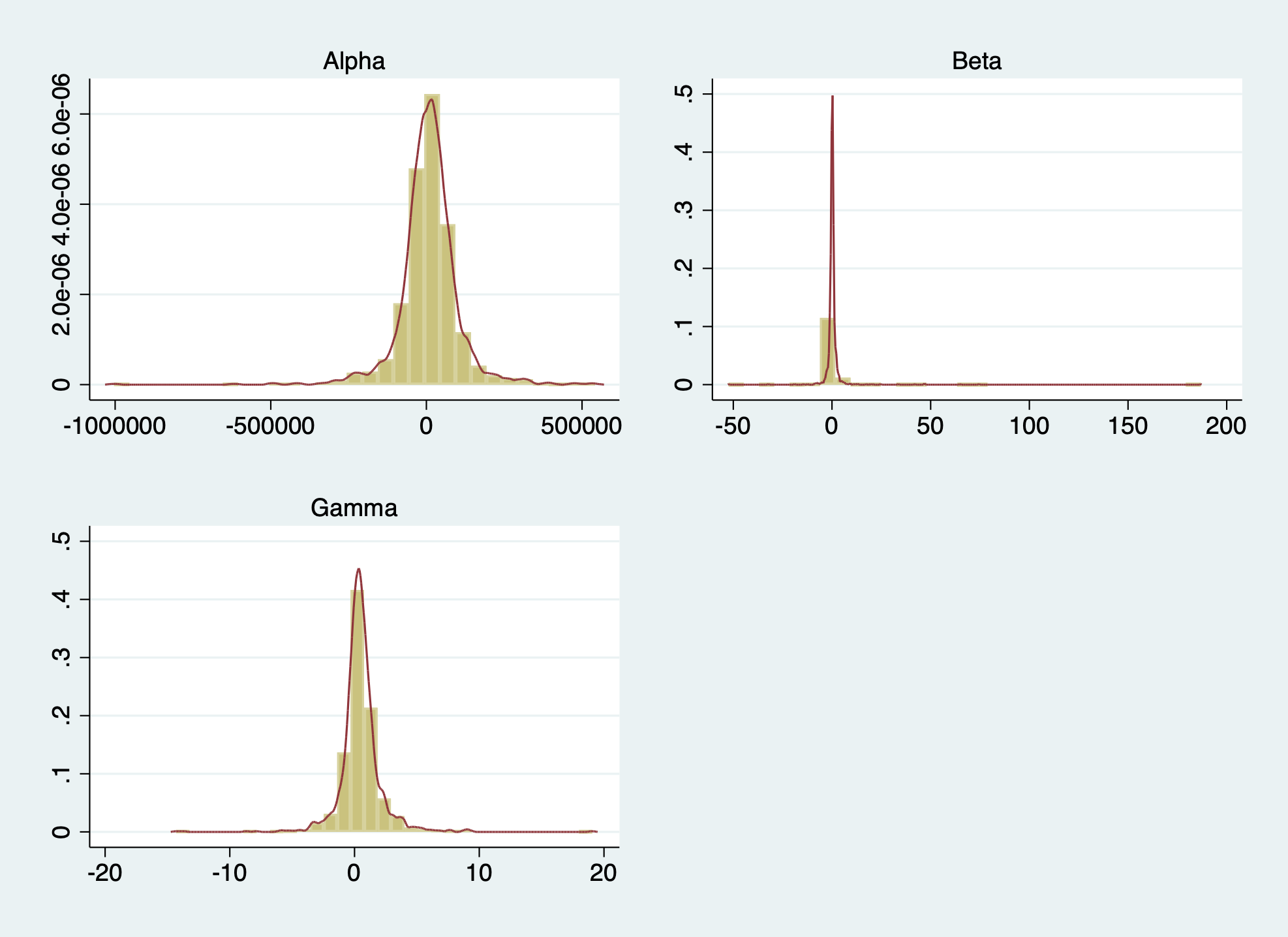}
%\vspace{-3.5cm}
\caption*{Colombia}\label{fig:ColombiaOLS}
\end{minipage}

\begin{minipage}[b]{.49\textwidth}
\centering
\vspace{1cm}
\includegraphics[width=.99\textwidth]{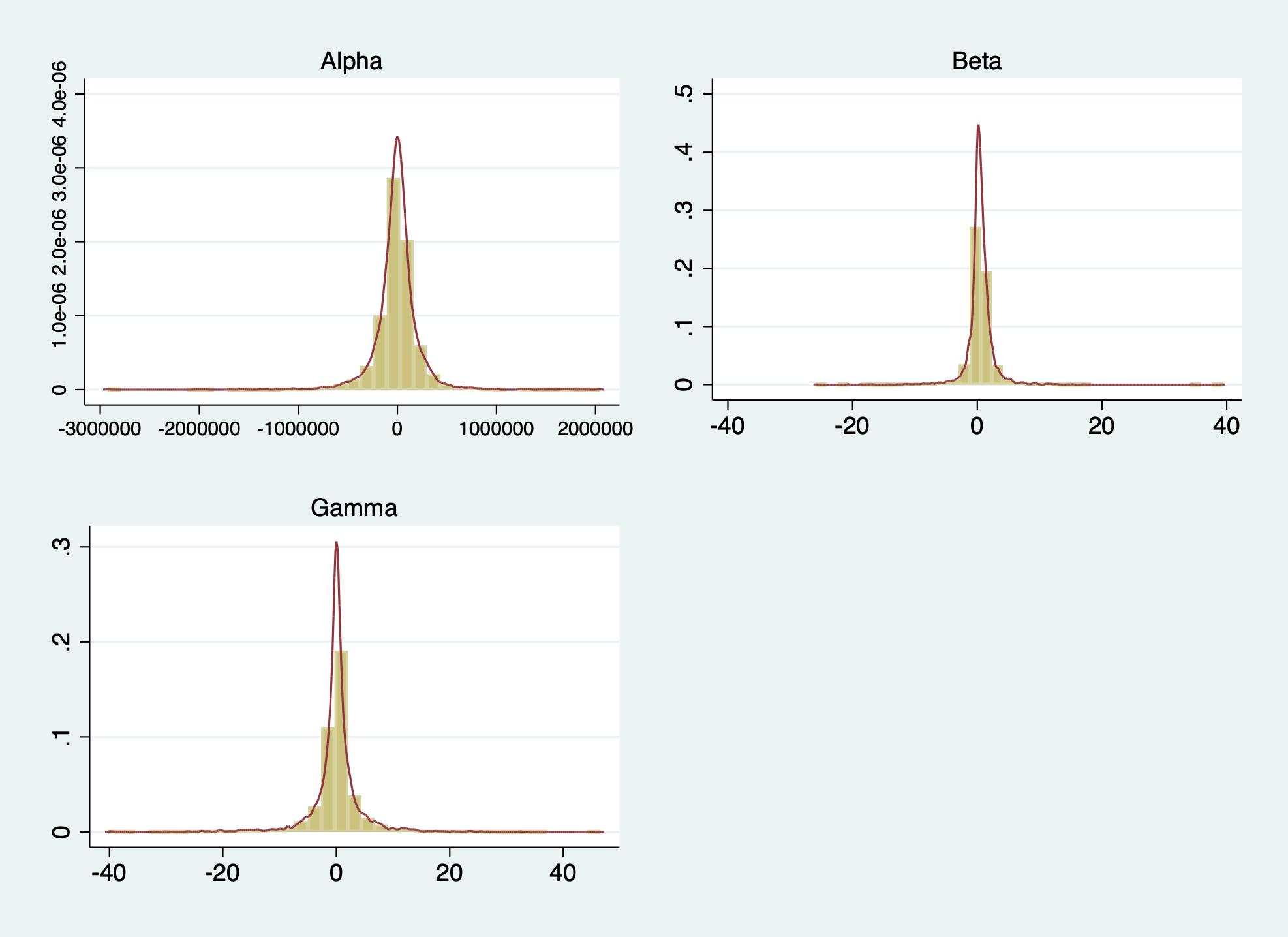}
%\vspace{-3.5cm}
\caption*{Japan}\label{fig:JapanOLS}
\end{minipage}

\caption{OLS CD coefficients distributions}\label{CD_OLS}

\end{figure}

\begin{figure}[H]

\centering
%\vspace{-2cm}
\begin{minipage}[b]{.49\textwidth}
\centering
\includegraphics[width=.99\textwidth]{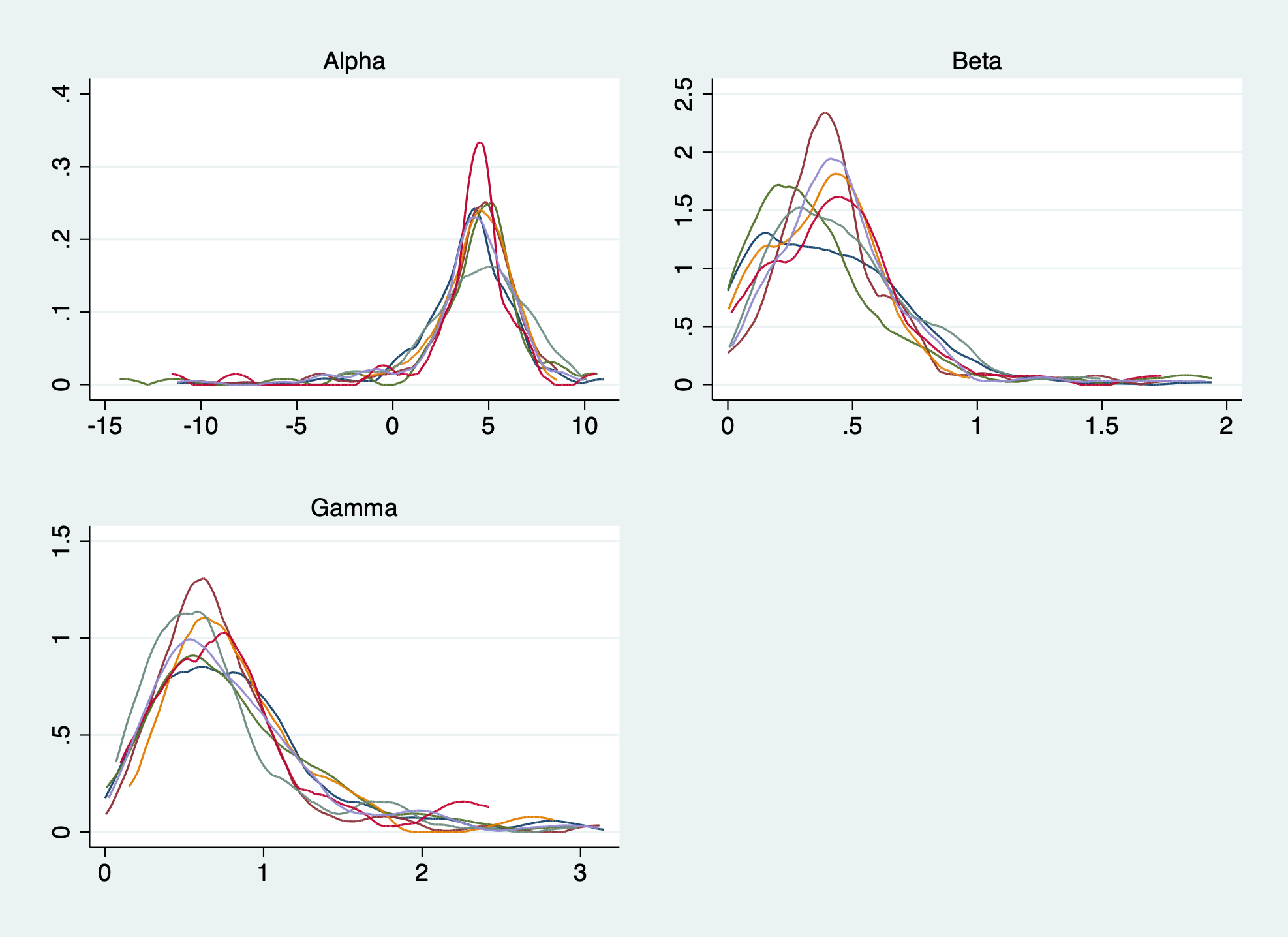}
%\vspace{-3.5cm}
\caption*{Chile}\label{fig:Chile by sectors}
\end{minipage}
\begin{minipage}[b]{.49\textwidth}
\centering
\includegraphics[width=.99\textwidth]{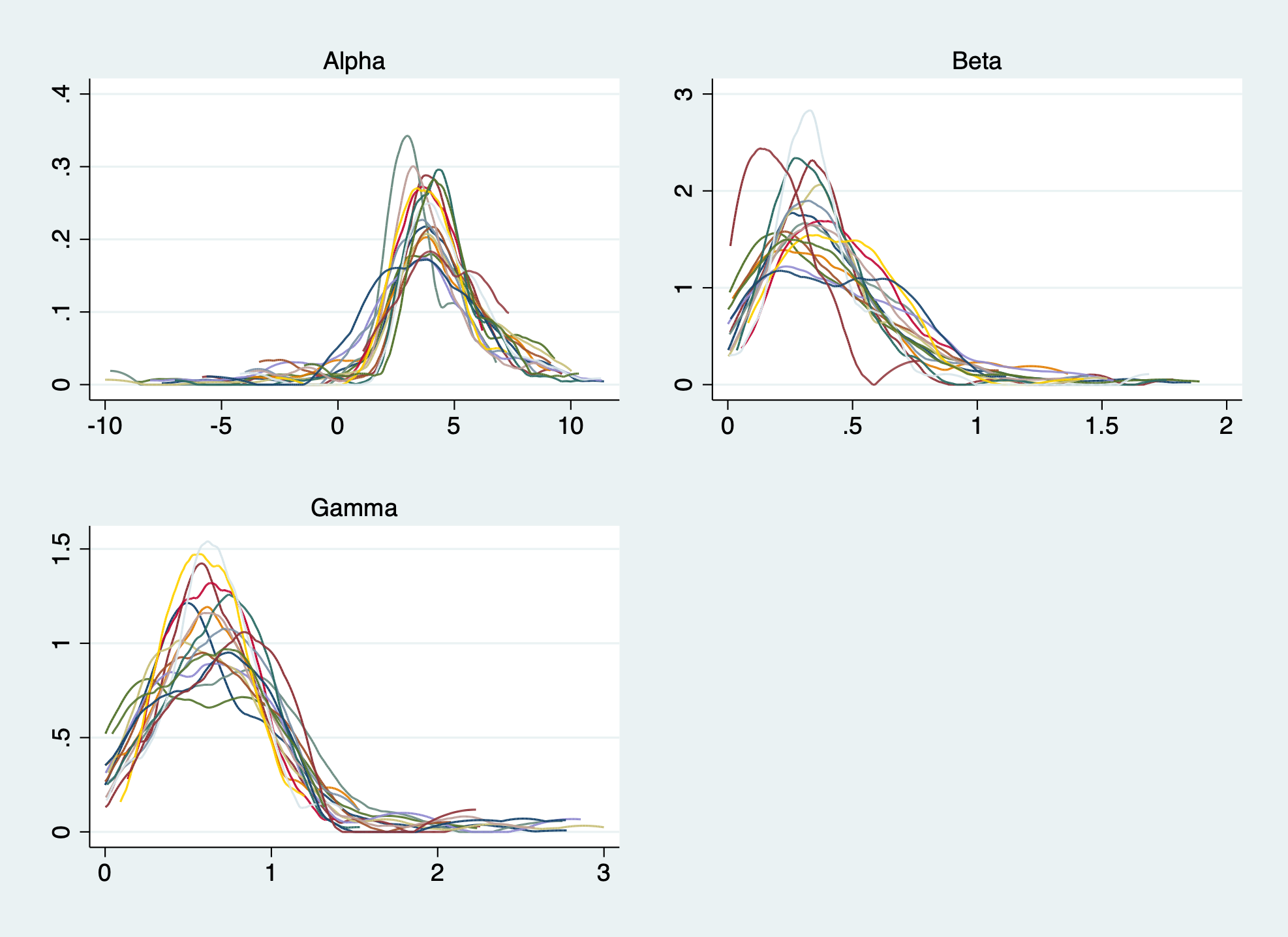}
%\vspace{-3.5cm}
\caption*{Colombia }\label{fig:Colombia by sectors}
\end{minipage}
%\vspace{0.5cm}
\begin{minipage}[b]{.49\textwidth}
\centering
\vspace{1cm}
\includegraphics[width=.99\textwidth]{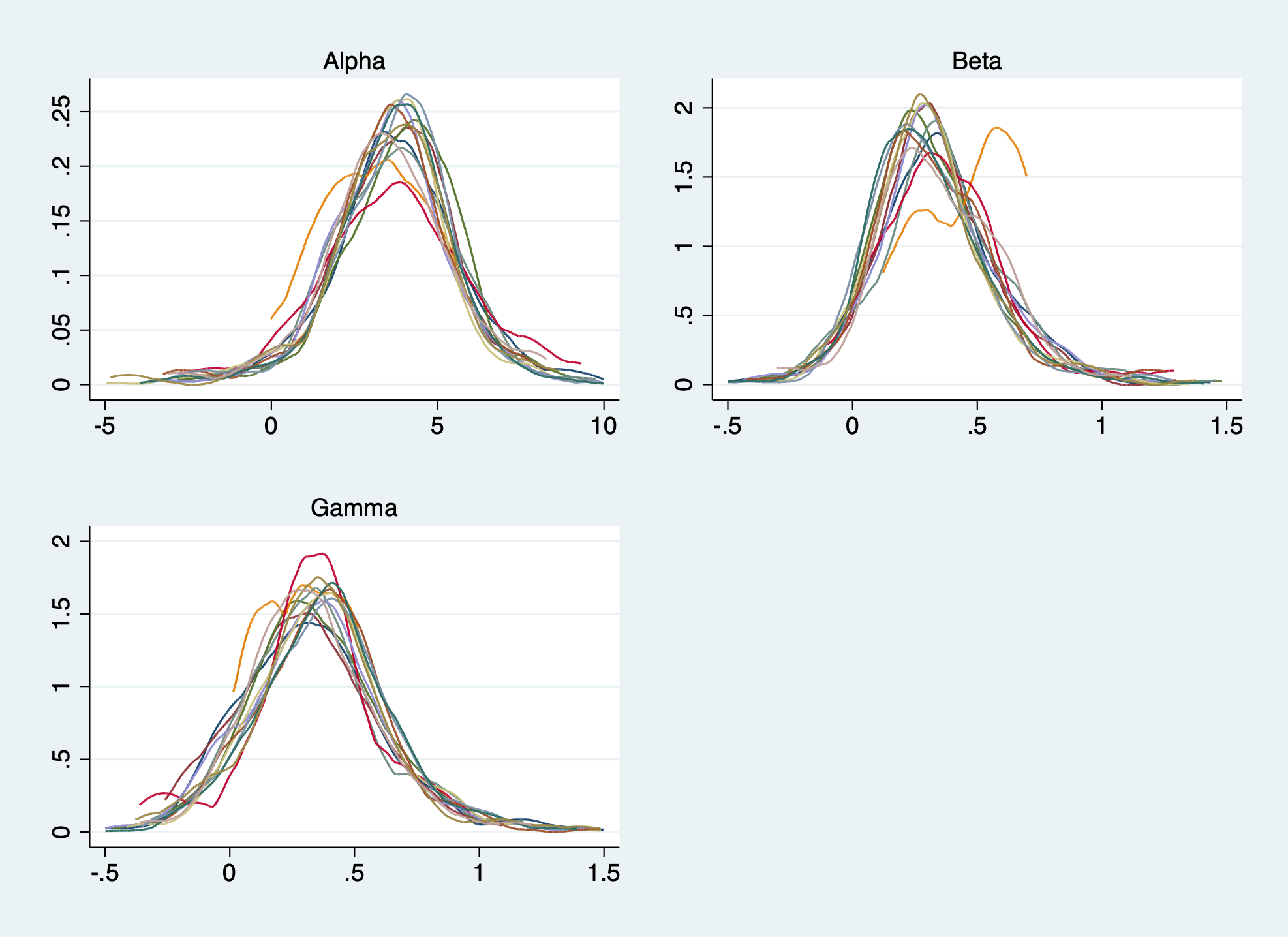}
%\vspace{-3.5cm}
\caption*{Japan}\label{fig:Japan by sectors}
\end{minipage}

\caption{CD coefficients distributions by sectors}\label{CD_nace}

\end{figure}

%%%%%%%%%%%%%%%%%%%%%%%%%%%%%%%%%%%%%%%%%%%%%%%%%%%%%

\newpage

\begin{figure}[H]

\centering
%\vspace{-2cm}
\begin{minipage}[b]{.49\textwidth}
\centering
\includegraphics[width=.99\textwidth]{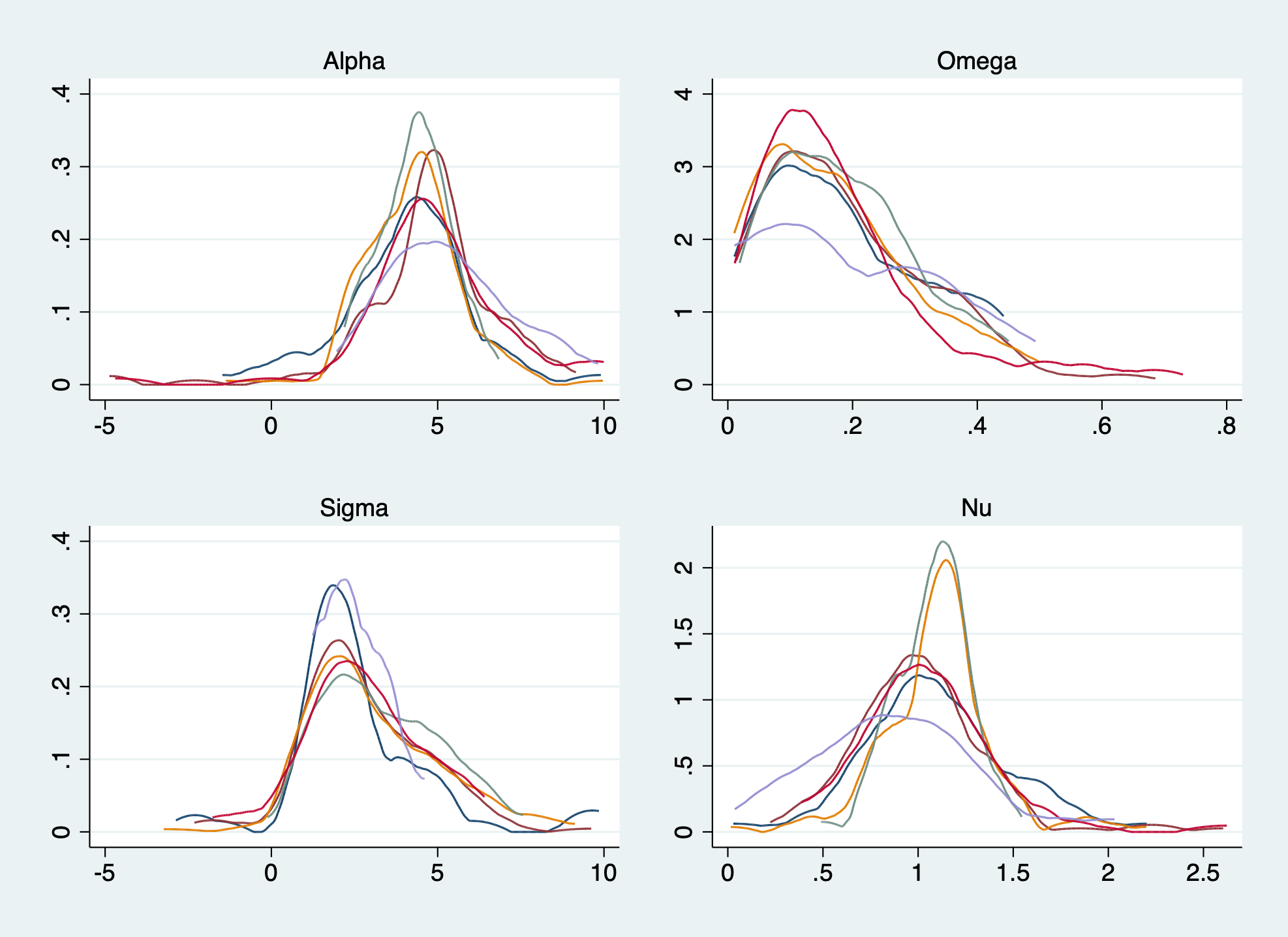}
\caption*{Chile}\label{fig:Chile by sectors}
\end{minipage}
\begin{minipage}[b]{.49\textwidth}
\centering
\includegraphics[width=.99\textwidth]{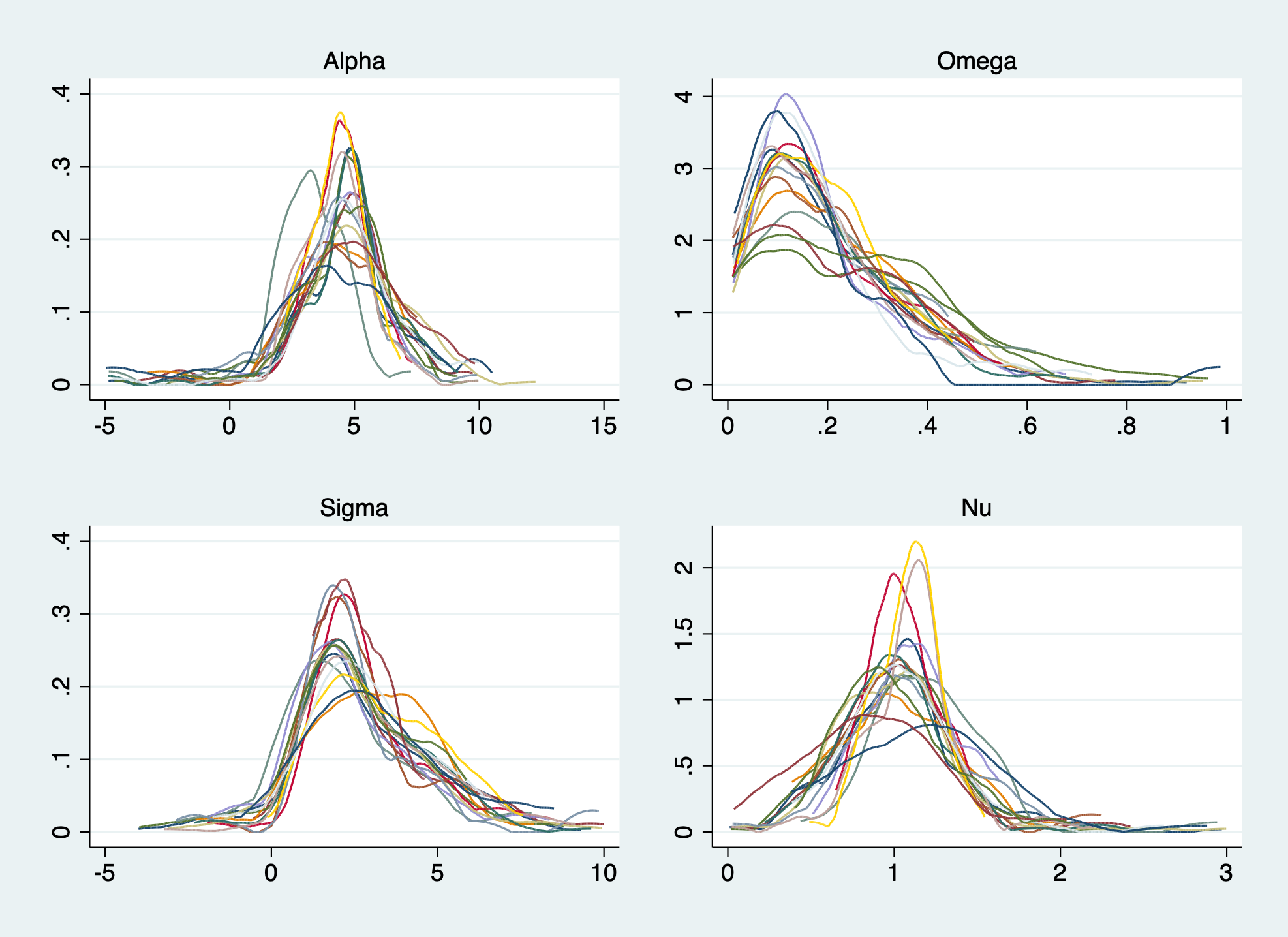}
\caption*{Colombia }\label{fig:Colombia by sectors}
\end{minipage}
\begin{minipage}[b]{.49\textwidth}
\centering
\vspace{1cm}
\includegraphics[width=.99\textwidth]{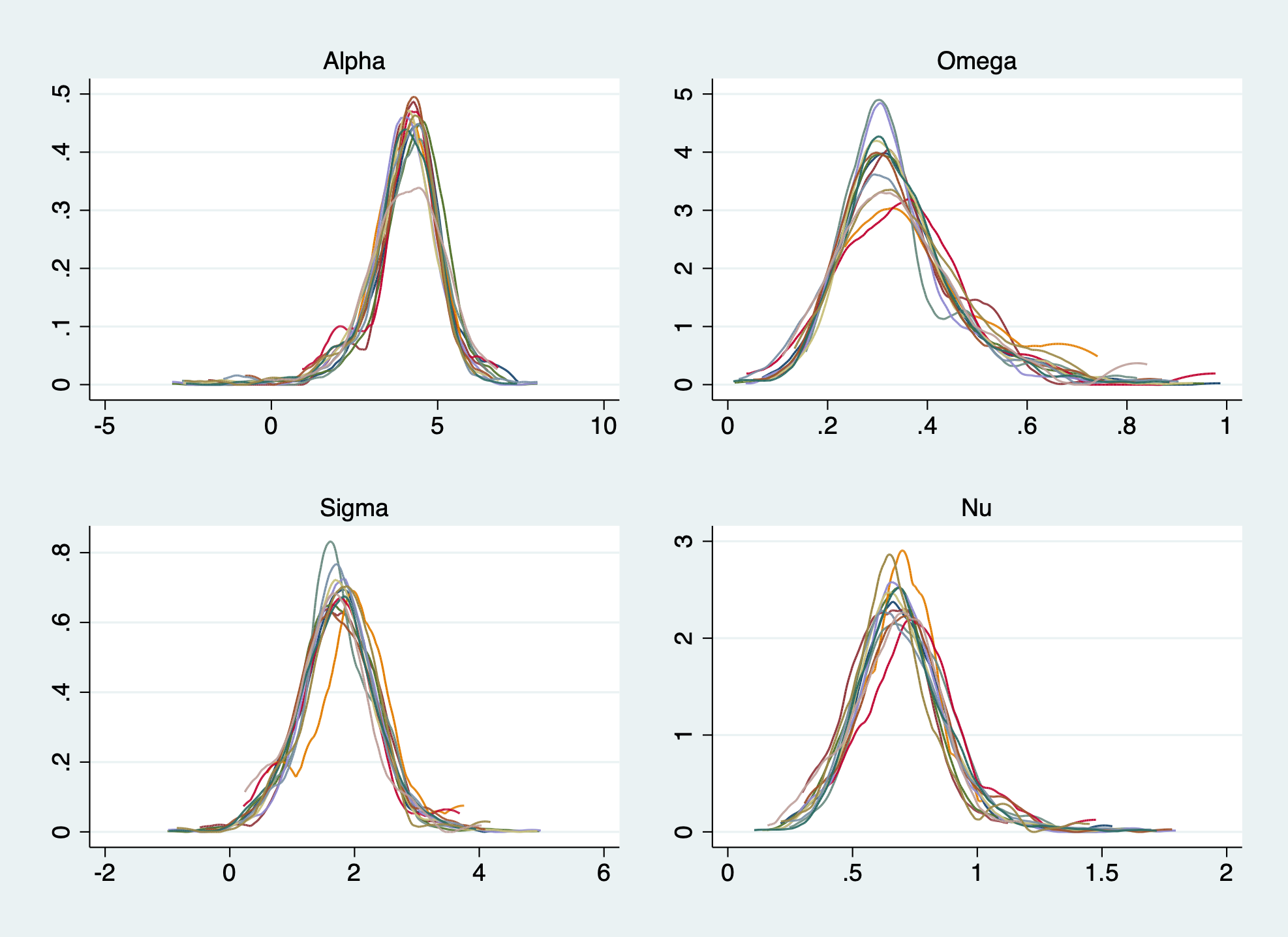}

\caption*{Japan}\label{fig:Japan by sectors}
\end{minipage}

\caption{CES coefficients distributions by sectors}\label{CES_nace}

\end{figure}

\end{document}